\documentclass[traditabstract]{aa}

\newcommand{\HII}{H\,{\scriptsize II}}
\newcommand{\hii}{H\,{\scriptsize II}}

\newcommand{\CII}{[C\,{\scriptsize II}]}

\newcommand{\OI}{[O\,{\scriptsize I}]}

\usepackage{epsfig,graphicx,color}
\usepackage{lscape}
\usepackage{ctable}

\usepackage{txfonts}
\usepackage{natbib}
\bibpunct{(}{)}{;}{a}{}{,}
\begin{document}

\bibliographystyle{aa}

\title{Globules and pillars in Cygnus X}
\subtitle{II. Massive star formation in the globule IRAS~20319+3958
\thanks{Based on observations from the Centro Astron\'{o}mico Hispano Alem\'{a}n (CAHA) at
Calar Alto, the Nordic Optical Telescope, La Palma, and the IAC80 telescope, Tenerife.}
}
  \author{A.A.Djupvik \inst{1}
  \and F. Comer\'on \inst{2}
  \and N. Schneider \inst{3}
}
 \institute{
  Nordic Optical Telescope, Rambla Jos\'{e} Ana Fern\'{a}ndez P\'{e}rez, 7, 38711 Bre\~{n}a Baja, Spain \\
  \email{amanda@not.iac.es}
  \and
  European Southern Observatory, Alonso de C\'ordova 3107, Vitacura, Santiago, Chile
  \and
  KOSMA, I. Physik. Institut, Z\"ulpicher Str.77, Universit\"at K\"oln, 50937 K\"oln, Germany
  }
%
%
\date{Received 19 May 2016 / Accepted 9 November 2016}
%

\abstract {Globules and pillars, impressively revealed by the
  {\it Spitzer} and {\it Herschel} satellites, for example, are pervasive features
  found in regions of massive star formation. Studying their embedded
  stellar populations can provide an excellent laboratory to test
  theories of triggered star formation and the features that it may
  imprint on the stellar aggregates resulting from it. We studied the
  globule IRAS~20319+3958 in Cygnus X by means of visible and
  near-infrared imaging and spectroscopy, complemented with
  mid-infrared {\it Spitzer}/IRAC imaging, in order to obtain a census
  of its stellar content and the nature of its embedded sources. Our
  observations show that the globule contains an embedded aggregate of
  about 30 very young ($\lesssim 1$~Myr) stellar objects, for which we
  estimate a total mass of $\sim 90$~M$_\odot$. The most massive
  members are three systems containing early B-type stars. Two of them
  most likely produced very compact \HII\ regions, one of them being
  still highly embedded and coinciding with a peak seen in emission
  lines characterising the photon dominated region (PDR). Two of these
  three systems are resolved binaries, and one of those contains a
  visible Herbig Be star. An approximate derivation of the mass
  function of the members of the aggregate gives hints of a slope at
  high masses shallower than the classical Salpeter slope, and a peak
  of the mass distribution at a mass higher than that at which the
  widely adopted log-normal initial mass function peaks.  The emission
  distribution of H$_2$ and Br$\gamma$, tracing the PDR and the
  ionised gas phase, respectively, suggests that molecular gas is
  distributed as a shell around the embedded aggregate, filled with
  centrally-condensed ionised gas. Both, the morphology and the low
  excitation of the \HII\ region, indicate that the sources of
  ionisation are the B stars of the embedded aggregate, rather than
  the external UV field caused by the O stars of Cygnus~OB2. The youth
  of the embedded cluster, combined with the isolation of the globule,
  suggests that star formation in the globule was triggered by the
  passage of the ionisation front.  }


\maketitle

\section{Introduction} \label{intro}

Pillars and globules are probes of the dramatic physical and structural changes
produced as a result of the erosion of the neutral gas at the interface between
\hii-regions and molecular clouds by the ionising radiation from OB-stars.
Expanding ionisation fronts encounter pre-existing dense condensations in the
molecular medium and overrun them, leaving them either isolated inside the
\hii~region (globules) or connected to the outside by a bridge of molecular gas
(pillars).  Recent simulations by \citet{tre12} show that the curvature of the
cloud surface plays a decisive role in this process (destruction or formation of
a pillar/globule).

\begin{figure*}[ht]
\begin{center}
\hspace{-0.5cm}
\includegraphics [width=14cm, angle={0}]{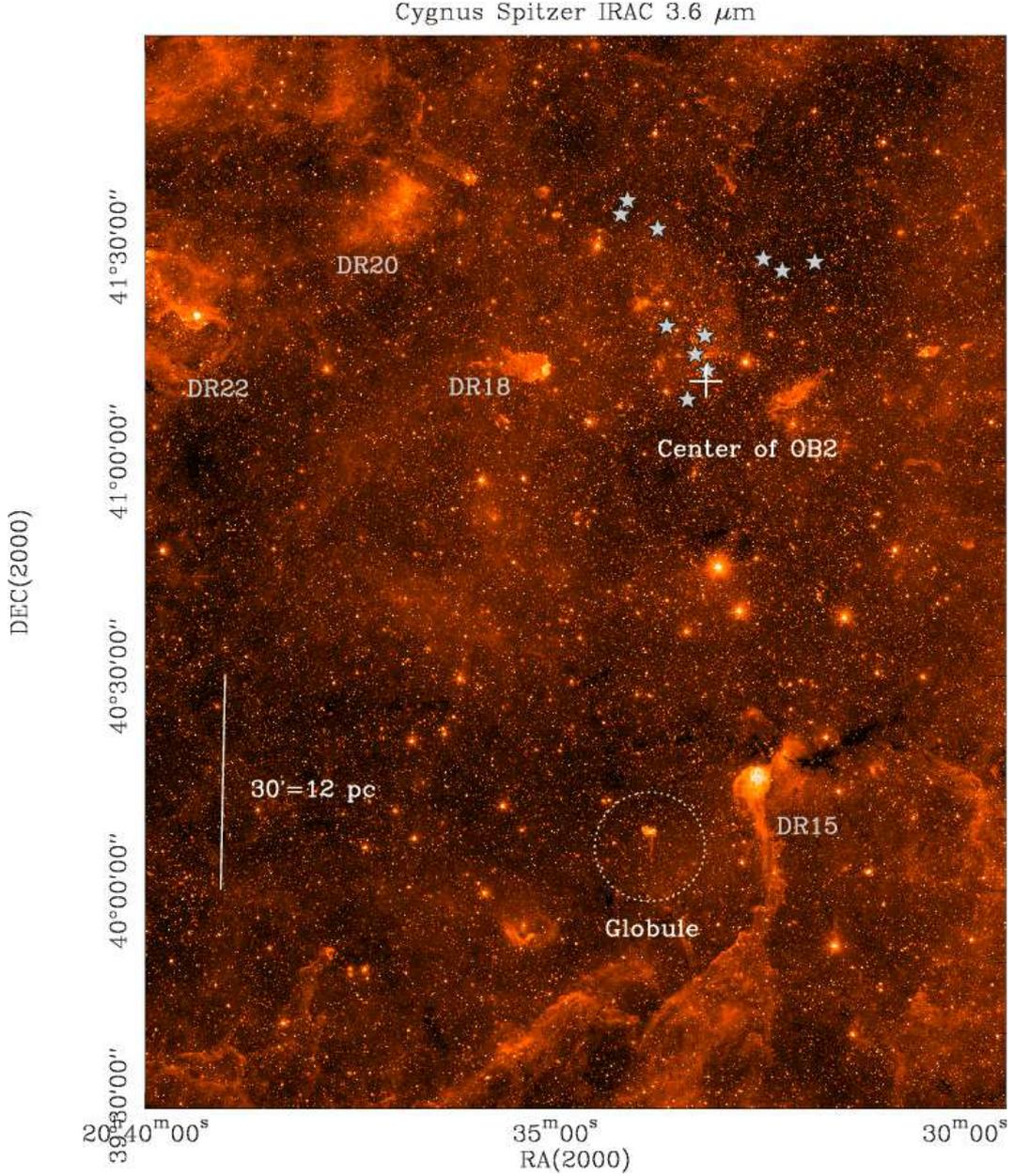}
\caption []{{\it Spitzer}\protect\footnotemark\  IRAC 3.6
  $\mu$m image of the central Cygnus X region including the globule
  (indicated by a dashed circle and located at $\alpha_{2000}$ =
  20$^h$33$^m$49.3$^s$ and $\delta_{2000}$ = 40$^\circ$08$'$52$''$).
  The most massive stars of the Cyg OB2 association are indicated with
  grey stars and the approximate center \citep[e.g.][]{wri15} with a
  white cross. The \hii\ regions DR15, 18, 20, and 22 are indicated.}
\label{overview}
\end{center}
\end{figure*}

Observations of some of these globules and pillars have shown them to
be active star-forming sites with bright embedded sources
\citep{com99,sug02}. Recent {\it Spitzer} images of the Cygnus~X
region (cf. Fig.~\ref{overview}), a vast molecular complex with
intense star-forming activity (see \citet{rei08} for an overview),
display abundant examples of globules and pillars formed as the
molecular cloud is progressively destroyed by the O stars of the
nearby Cygnus OB2 association
\citep[e.g.][]{kno00,com02,com12,wri14}. Detailed observations of
far-infrared lines of \CII , \OI\ and CO J=11$\to$10 observed with
SOFIA\footnote{Stratospheric Observatory for Far Infrared Astronomy,
  https://www.sofia.usra.edu/} \citep{sch12} and {\it Herschel}
\citep{sch16}, have revealed the peculiar nature of one globule
  in Cygnus X, \object{IRAS~20319+3958} (hereafter referred to as 'the
  globule'). The SOFIA study indicates that the globule is highly
  dynamic, with velocity features corresponding to rotation and
  outflowing gas. With the {\sl Herschel} data, the physical
  properties of the globule were determined, giving a mass of
  $\sim$240 M$_\odot$ and a temperature of $\sim$20 K. Most
  interestingly, however, is the finding that the observed UV-flux is
  of the order of 550 G$_\circ$ (with G$_{\circ}$ in units of the Habing 
  field 2.7$\times$10$^{-3}$erg cm$^{-2}$ s$^{-1}$), much
  higher than that found for all other globules in the Cygnus X
  region.

\footnotetext{Data taken within the Cygnus-X legacy survey, see
https://www.cfa.harvard.edu/cygnusX/.}

The globule hosts a small, moderately embedded cluster. \citet{kro06} and \citet{kum06}
reported nine cluster members within a projected radius of 0.53 pc, with an estimated
stellar mass of 17 M$_\odot$. Here we adopt a distance of 1.4 kpc for the globule, which
is the one determined for structures associated with the Cygnus~X complex \citep{ryg12}
from parallax observations. A previous study by \citet{coh89}, motivated by mid-infrared
emission features ascribed to PAHs, led to the identification of two visible stars as
the source of excitation, for which those authors estimated mid-B spectral types.

A question not answered with the observations presented in \citet{sch12} is whether the 
current physical conditions of the gas in the globule are mainly determined by its external 
illumination by the Cyg OB2 cluster, or whether the brightest stars of the
\object{IRAS~20319+3958} cluster produce their own photon dominated region (PDR) and \hii\ 
region. One of the objectives of this paper is to answer this question and to reveal the 
stellar content of the globule. Previous studies \citep{sch12,sch16} have proposed that
the globule is one of the rare examples of massive/intermediate-mass star formation in 
globules found in Cygnus~X. Our finding is that the globule hosts a group of more than 30 
very young stars, of which at least two are early B-type stars and likely the origin of the 
ionised gas found in the centre of the stellar aggregate.

For that, we obtained and present here deep near-infrared imaging of the globule complemented
with visible imaging, and with spectroscopy in the visible and infrared of selected cluster
members. The combination of near-infrared and existing {\it Spitzer} observations allows us
to obtain a much more complete census of cluster members than previously available and to study
in detail the evolutionary state of the cluster and some of its most outstanding members.
At the same time, the detailed views of the respective distributions of the PDR and the ionised
gas, together with the diagnostics provided by selected emission lines, yield further insight
into the interaction of the cluster members with the various phases of the interstellar gas in
the globule.

This paper (the second one of a series) is part of a larger study on understanding pillar and
globule formation, and the conditions under which stars can form within them (PI: N. Schneider).
The project draws together far-infrared imaging and spectroscopy from {\it Herschel},
{\it Spitzer}, and SOFIA, and ground-based millimeter molecular-line
mapping\footnote{https://hera.ph1.uni-koeln.de/$\sim$nschneid/}.


\section{Observations and data reduction} \label{obs}

\subsection{Near-IR imaging}

\subsubsection{Omega2000 observations}

The $J$, $H$, $K_S$, Br$_{\gamma}$ and H$_2$ imaging was obtained in service mode with the Omega2000 
infrared camera on the 3.5m telescope in Calar Alto on 29th and 30th of June, 2012. The Hawaii-2 
array (2048 $\times$ 2048 $\times$ 18 $\mu$m) with a pixel resolution of 0$\farcs$449 gives a total 
field-of-view 
(FoV) of 15.4' $\times$ 15.4'. The imaging was done with a dither pattern consisting of 15 positions, 
using an integration time of approximately 1 minute on each position. The total integration time was 
15 minutes for the broad-band filters $J$, $H$ and $K_S$ and 1 hour for the Br$_{\gamma}$ and H$_2$ 
narrowband filters. Details are given in Table~\ref{tab-obs}.

The images have been reduced using the IRAF package and a set of own scripts. Bad pixels were treated 
in all images by interpolation, using a bad pixel mask provided by the observatory. Flat field correction
was made using differential (lamp ON-OFF) dome flats for each filter. For each filter, 15 dithered 
images, obtained within approximately 20 minutes, were scaled and median filtered to make a sky image 
for subtraction, after re-scaling. The images were shifted and combined. The plate solution for each 
final image was found using the 2MASS catalogue \citep{skr06} as a reference, and all images were placed on 
the same scale.

\begin{table}[t]
\caption{Omega2000 observations. Each of the N$_{\rm im}$ dithered images is a co-addition of N$_{\rm coadd}$ 
individual integrations of T$_{\rm dit}$ single exposure time.}

\begin{tabular}{lcrrrrrrr}
\hline
Band & $\lambda_c$ & T$_{tot}$ & N$_{im}$ & T$_{dit}$ & N$_{coadd}$ &
FWHM & Airmass \\
\hline
      & ($\mu$m)    & (sec)     &          & (sec)     &             & ('') & \\
\hline
$J$      & 1.209 & 900 & 15 &  6 &   10 &  2.0 &  1.155 \\
$H$      & 1.647 & 900 & 15 &  3 &   20 &  1.6 &  1.110 \\
$K_S$  & 2.151 & 900 & 15 &  2 &   30 &  1.7 &  1.076 \\
Br$_{\gamma}$ &  2.166 & 3600 &  60 &  20 & 3 & 1.2 &  1.007 \\
H$_2$  & 2.122 & 3600 &  60  &  20  &  3 & 1.8 &  1.412 \\
\hline
\end{tabular}
\label{tab-obs}
\end{table}

Due to the chosen dither pattern, the region of interest for our study has an uneven coverage. The deepest 
part is a combination of 15 images while the southernmost area is made from only 3 images, and therefore 
the noise varies spatially across the image. Point source detection was done subregion by subregion using 
{\it daofind} in the IRAF {\it digiphot} package and with a threshold of 5 sigma and a conservative limit
on $\sigma_{sky}$ in each subregion. For this reason our estimate of completeness is uncertain. PSF 
photometry was made on all sources in each filter, using a penny1 model of the PSF that varies 
quadratically over the FoV. A total of 12921 stars were measured in the $H$-band which results in the 
deepest image. Using the plate solution based on 2MASS positions, RA/DEC coordinates were calculated for 
all sources and the table was cross-correlated with 2MASS, finding a total of 2215 stars in common. The 
derived positions of the Omega2000 sources are estimated to be accurate to approximately 0$\farcs$1. For 
the photometric calibration, we use 2MASS stars with quality flag `AAA' (837 stars). The offsets between 
Omega2000 and 2MASS magnitudes were examined for trends in colours, as well as spatial trends over the 
detector. There is a clear dependence on Y position for the $J$-band, and we used an empirical correction 
as a function of Y-pixel by a linear fit to the data ($Jc = J - 4.9 \times 10^{-5} \times$~Y-pixel). In $H$ 
and $K_S$ no clear spatial dependence was found. Colour trends were then checked by plotting magnitude and 
colour differences versus the $(J-H)_{\rm 2MASS}$ colour index over a range from 0 to 3.3 magnitudes, giving 
the following results: 
{\small
\begin{eqnarray}
\Delta (H-K_S) = 0.004 \pm 0.007 - 0.001 \pm 0.005 \times (J-H)_{2MASS} \\
\Delta (J-H) = -0.026 \pm 0.006 + 0.020 \pm 0.005 \times(J-H)_{2MASS},
\end{eqnarray}
}
where $\Delta$ indicates the difference (Omega2000 - 2MASS) for the index in question. We found these 
sufficiently small to safely neglect any colour corrections, since neither the foreground nor the internal 
extinction cause strong enough reddening on the sources of interest. The median offsets between 2MASS and 
Omega2000 magnitudes were then used to calibrate our $JHK_S$ magnitudes. The scatter in the offsets are 
large, however, as expected due to different spatial resolution. In order to estimate the accuracy of the 
calibration we select sources with $K_S < 12$~mag that have no IR excess and are not blended with other 
sources, limiting the number to approximately 140 stars. On this sample we use a 2-sigma clipping and find 
standard deviations around the mean offset of 0.043, 0.024 and 0.037~mag in $J$, $H$ and $K_S$, 
respectively. We take this as an estimate of our calibration error and add it up in quadrature to the 
output error from the PSF photometry.

In the source table, we have excluded Omega2000 magnitudes for sources brighter than approximately 10.4, 
10.0 and 8.8 mag at $J$, $H$ and $K_S$, respectively, since these bright stars enter the non-linear range 
of the Omega2000 array (25000 ADU) for our choices of individual exposure times. This means that in our 
further analysis, we use the 2MASS magnitudes and, where available, NOTCam magnitudes for these sources.
In the deepest part of the field, the $5\sigma$ magnitude limit, defined as S/N=5 in the final photometry, 
is reached at J = 20.5, H = 19.7, and K$_S$ = 18.5 mag. We make an approximate and conservative estimate 
of the completeness to be 18.5, 17.5 and 16.5 mag in $J$, $H$ and $K_S$, respectively, based on the 
detection histograms, but note that the observations have a different depth across the field, making such 
an estimate uncertain.

\begin{figure}[t]
\begin{center}
\includegraphics[angle=0,width=9cm]{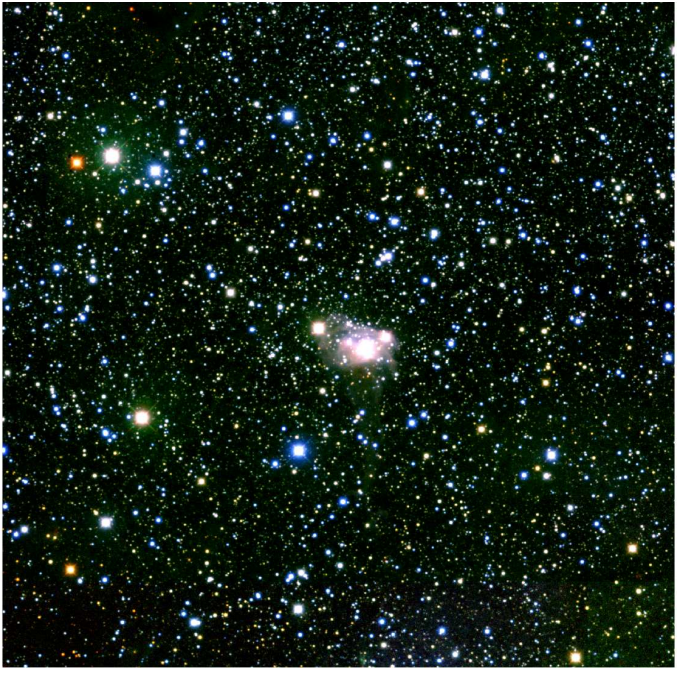}
\caption [] {JHK$_S$ RGB combined image of a 15$'\times$15$'$ field centred on the globule at 
$\alpha_{2000}$ = 20$^h$33$^m$49.3$^s$ and $\delta_{2000}$ = 40$^\circ$08$'$52$''$ obtained with Omega2000. 
North up and east left.}
\label{o2kmap}
\end{center}
\end{figure}

\subsubsection{NOTCam observations}

At the Nordic Optical Telescope (NOT), La Palma, the NOT's near-IR camera and spectrograph NOTCam 
\citep{abb00,dju10} was used in its high-resolution mode, with a pixel scale of 0$\farcs$078/pix and a FoV 
of 80$\arcsec \times 80\arcsec$, on 6th September, 2012, to image the central part of the globule in 
photometric and good seeing conditions. To avoid saturation, we used 3.6 seconds individual integration 
time and dithered the telescope in a 3x3 grid. Discarding a few individual images in the reductions, we 
obtained total integration times of 21.6, 25.2 and 28.8 seconds in $J$, $H$ and $K_S$, respectively.

All raw images were first corrected for non-linearity using the pixel-by-pixel correction coefficients 
available from the NOTCam web pages\footnote{For details on the calibration of NOTCam see 
http://www.not.iac.es/instruments/notcam/}. After this the images were bad-pixel corrected, flat-field 
corrected using differential skyflats, and sky-subtracted using the frames themselves to estimate the sky. 
The shifted images were combined, resulting in a FWHM of the PSF in the combined images of approximately 
0$\farcs$4.

PSF fitting photometry was made on the point sources detected, and magnitudes were found for 177 stars in 
$H$ and $K$ and 64 stars in $J$. Selecting stars with quality flag `AAA' from the 2MASS catalogue, 
excluding multiples resolved by NOTCam, results in nine stars that are useful to calibrate magnitudes and 
positions. The positional accuracy with respect to 2MASS positions is 0$\farcs$1, and the standard 
deviation of the difference magnitudes m$_{\rm NOT}$ - m$_{\rm 2MASS}$ is 0.08, 0.06 and 0.06 mag in $J$, $H$ 
and $K_S$, respectively. This calibration error is added in quadrature to the individual errors in the PSF 
magnitudes. The completeness limit is estimated to be 18, 18 and 17 mag for $J$, $H$ and $K_S$, 
respectively.

\subsection{Visible imaging}

Images centered on \object{IRAS~20319+3958} were obtained through the $VR_CI_C$ filters using the CAMELOT 
camera at the IAC80 telescope at the Teide Observatory on the night of 20th June, 2012. The 
camera covers a FoV of 10$\farcm$1 $\times$ 10$\farcm$1 on the sky. The exposure times were 600s, 300s 
and 120s through the $V$, $R_C$ and $I_C$ filters, respectively. Flux calibration was achieved through 
observations of the nearby cluster NGC~6910 in all three filters with exposure times of 60s ($V$), 30s 
($R_C$) and 10s ($I_C$) when the cluster was near meridian transit, followed by observations with the same 
exposure times centered on the globule. The deeper images were obtained immediately before those of 
NGC~6910, thus ensuring that all observations were taken at virtually the same air mass. Standard image 
data reduction was carried out using IRAF, and PSF instrumental photometry on the resulting images was 
carried out with dedicated IRAF scripts using the layered DAOPHOT package \citep{ste87}. Astrometry was 
carried out using stars in common with the UCAC4 catalogue \citep{zac13} and used to cross-match the sources 
detected in the CAMELOT images with those detected in the infrared observations.

Since no dedicated $VR_CI_C$ photometry of NGC~6910 seems to exist in the literature, the CCD $Vr'i'$ 
photometry of stars in the field of NGC~6910 included in the UCAC4 catalogue was used to determine 
photometric zero-points in each filter using the transformation equations between the $Vr'i'$ and the 
$VR_CI_C$ photometric systems given by \citet{fuk96}:
\begin{eqnarray}
R_C = r'-0.19(V-r')-0.15 ,\\
I_C = R_C - 0.23 - 1.02 (r'-i') 
.\end{eqnarray}

We then transferred the zero-points obtained on the NGC~6910 field to the shallow exposures of the globule 
field. We preferred this two-step approach rather than the direct calibration using the UCAC4 $Vr'i'$ 
photometry of stars in the globule field because the former field is more crowded, thus allowing us to 
obtain a more robust determination of the photometric zero-points. The estimated $5\sigma$ detection limits 
of the deep exposures of the globule are $V \simeq 20.0$, $R_C \simeq 19.5$, and $I_C \simeq 18.0$. 
Saturation in the deep images was reached at $V \simeq 15.0$, $R_C \simeq 14.0$, and $I_C \simeq 14.0$. 
The photometry of the stars brighter than these limits was taken from the shallow images of the field used 
for photometric calibration.

\subsection{Optical spectroscopy}

A set of optical spectra of the two brightest stars (referred to as star~$A$ and star~$B$ in 
Sect.~\ref{results}) that appear projected on the globule's head at visible wavelengths was obtained on the 
night of 25th August, 2012, using the TWIN spectrograph at the Calar Alto 3.6m telescope. A dichroic mirror 
was used allowing us the simultaneous use of the blue and red arms of TWIN, resulting in a continuous 
spectral coverage in the range 4000~\AA \ to 7500~\AA. The gratings used yielded a resolving power of 
$R = 2400$ in both the blue and red arms, using a common entrance slit of 1$\farcs$2 width. The slit was 
oriented at a fixed position angle so as to include both stars in a single setting.
The slit length, of 6$'$, greatly exceeds the 30$''$ of separation between those stars. Slit losses due to 
differential refraction are minimal due to the fact that the observations were obtained when the target 
region was near meridian transit, a few degrees from the zenith, at a maximum air mass never exceeding 1.12.
Eight sets of spectra with an exposure time of 1800s each were obtained in each arm. The individual science 
frames were bias-subtracted and divided by a flat field obtained with the telescope dome illuminated 
by a continuum lamp the morning after the observations. Wavelength calibration frames using a thorium-argon 
lamp were obtained immediately before and after the set of observations.

One-dimensional spectra of each of the two bright stars were extracted from each frame using a fixed-width 
aperture centered on the spectral trace of each of them. In addition, a spectrum of the brightest portion 
of the \HII\ region, located approximately 9$''$ East of the bright star near the center of the globule's 
head, was also extracted. Background subtraction from the stellar spectra was carried out by averaging the 
spectrum extracted from two apertures adjacent to each star, and the same was done for the spectrum of the 
nebula. In this way, the spectrum of the all-pervading, structureless \HII\ nebulosity present all along the 
slit is subtracted from the extracted spectra of both stars and of the nebula.

\subsection{$K$-band spectroscopy}

$K$-band spectra were obtained with NOTCam for the three brightest young stellar object (YSO) candidates in 
the globule. Two different spectral resolutions were used, both applying grism~\#1 and the K-band filter 
\#208. In low-resolution mode the wide-field camera and the 128~$\mu$m slit (a long slit of width 0.6$''$ 
or 2.6 pixels) covers the range from 1.95 to 2.37 $\mu$m with a dispersion of 4.1 \AA /pix, giving a 
resolving power $R = \lambda /\Delta \lambda$ of approximately 2100. In medium-resolution mode the 
high-resolution camera lens together with the 44 $\mu$m slit (0.2$''$ or 2.6 pixels wide) covers the range 
from 2.07 to 2.20 $\mu$m with a dispersion of 1.4 \AA /pix and a resolving power $R \sim$ 5500. The spectra 
were obtained by dithering along the slit in an A-B-B-A pattern, reading out the array non-destructively 
while sampling up the ramp. Argon and halogen lamps were obtained while pointing to target in order to 
minimise effects due to instrument flexure and to improve fringe correction. To correct the spectra for 
atmospheric features we used HD199373 (F5 V) and HIP99346 (F0 V) as telluric standards.

\begin{table}[t]
\caption{NOTCam $K$-band spectra listed with date, target name, resolving power, exposure time, seeing 
conditions and spatial pixel resolution.}
\begin{tabular}{llrrrl}
\hline
Date        & Star & $\lambda /\Delta \lambda$ &  T$_{\rm exp}$ & FWHM & Pixel scale \\
\hline
22 Jul 2013 & $A$ &  2100    &  120s $\times$ 8 & 0.5$''$ & 0.234$''$ \\
02 Aug 2015 & $A$ &  5500    &  300s $\times$ 4 & 0.5$''$ & 0.078$''$  \\
05 Oct 2014 & $B$ &  2100    &   60s $\times$ 8 & 0.6$''$ & 0.234$''$  \\
22 Aug 2014 & $C$ &  2100    &  300s $\times$ 4 & 0.7$''$ & 0.234$''$  \\
\hline
\end{tabular}
\label{tab-kspec}
\end{table}

The 0$\farcs$5 separation double star~$A$ was first observed in good seeing in low-resolution mode
with the slit aligned to include both components, but it was difficult to separately extract the
two spectra. The medium-resolution spectrum, however, which has three-times higher spatial
resolution, resolved both components $A_N$ and $A_S$ well. The 1$\farcs$4 separation double star~$C$
(see Sect.~\ref{results}) was easily spatially resolved in the low-resolution mode, and star~$B$ is
apparently single. Slit positions are overlaid on a high-resolution NOTCam $K$-band image in
Fig.~\ref{slitimage}.

\subsection{Spitzer IRAC data}

Mid-IR photometry was obtained from the archive {\it Spitzer} post-BCD images in the four IRAC bands 
at 3.6, 4.5, 5.8 and 8.0 $\mu$m in the 15$'\times$15$'$ field mapped in JHK$_S$ with Omega2000. The 
point source photometry was done through the PSF fitting procedure as described in \citet{com11}. The 
flux-to-magnitude conversion was made using the fluxes for a zeroth magnitude star as given in the 
IRAC and MIPS web pages. The catalogue of IRAC photometry was cross-correlated with the Omega2000 
table using a tolerance of 1$\farcs$3 in the matching of RA and DEC coordinates. We find a median
offset of 0$\farcs$22 in RA and 0$\farcs$04 in DEC, and a total of 3476 sources are correlated 
(compared to 2119 correlated with 2MASS). The minimum magnitudes for point sources before entering 
saturation (for a frame-time of 12s) was found to be 9.8, 9.3, 6.7 and 6.8 mag for the bands 3.6, 4.5, 
5.8 and 8.0 $\mu$m, respectively. Sources brighter than this are flagged as upper magnitude limits. 
The 5$\sigma$ point source detection limits are found at 15.3 mag, 14.8 mag, 12.7 mag and 11.7 mag 
for the bands 3.6, 4.5, 5.8 and 8.0 $\mu$m, respectively.


\section{Results and analysis}
\label{results}

\subsection{YSO classification}
\label{irex}

We analyse the Omega2000 JHKs photometry and the {\it Spitzer}/IRAC photometry in a 15$'\times$15$'$
field centred on the globule at $\alpha_{2000}$ = 20$^h$33$^m$49.3$^s$ and $\delta_{2000}$ =
40$^\circ$08$'$52$''$ and shown in Fig.~\ref{o2kmap}. The IRAC photometry allows us to use well-established 
criteria to select Young Stellar Objects (YSOs) and classify them on the basis of their spectral energy 
distribution (SED) in the 3.6~$\mu$m to 8~$\mu$m region. Additional near-IR $JHK_S$ photometry extends the 
SEDs towards shorter wavelengths and helps characterising fainter YSOs that are not detected in all IRAC 
bands because of lower spatial resolution and sensitivity. The near to mid-IR photometric analysis is based 
on detecting IR excesses owing to circumstellar dust that absorbs and re-emits in the infrared. The slope 
of the SED from near to mid-IR, the $\alpha$ index, defined as the slope of $\log (\lambda F_{\lambda})$ vs 
$\log (\lambda)$, originally defined between 2 and 20 $\mu$m, has been used to define observed YSO classes 
and link them to the early stellar evolutionary stages \citep{lad84,ada87,and93,gre94}, see \citet{eva09}
for a review. Class~IIIs have $\alpha < -1.6$, Class~IIs occupy a region of $ -1.6 < \alpha < -0.3$,
flat-spectrum sources have $ -0.3 < \alpha < 0.3$, and Class~Is have $\alpha > 0.3$.  Class~0s are
often not directly detected at mid-IR wavelengths and are classified by their large submillimeter
luminosity. Class~I sources are embedded protostars with infalling dust envelopes, flat-spectrum
are thought to have cleared out a cavity in their dust envelope, and Class~IIs are pre-main sequence
(PMS) stars with dust disks. Class~IIIs have no detectable IR excesses and can not, therefore, be
distinguished from field stars with near- and mid-IR photometry alone. An alternative to the SED
slope is to use appropriately defined loci in colour-colour (CC) diagrams to distinguish between the
YSO classes and separate IR-excess sources from reddened field stars. For IRAC photometry in particular,
a number of slightly different colour criteria have been used by different authors
\citep{meg04,all04,gut08,gut09}.

\begin{figure}
\includegraphics[angle=270,width=9cm]{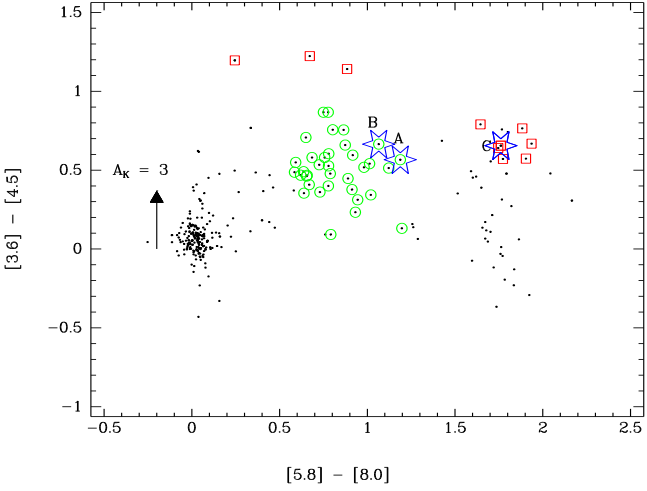}
\caption []{The IRAC $[3.6]-[4.5]/[5.8]-[8.0]$ diagram for the sample of 316 sources with photometric
errors $<$ 0.2 mag in all bands (black dots) and the selected Class~I (red squares) and Class~II
(green circles) candidates using the criteria given in the text. The location of the three
sources A, B and C are marked with blue star symbols. The direction and size of the reddening vector
for $A_K$ = 3 mag is according to empirical relations for the IRAC bands by \citet{ind05}.
}
\label{irac1234}
\end{figure}

\begin{figure}
\includegraphics[angle=270,width=9cm]{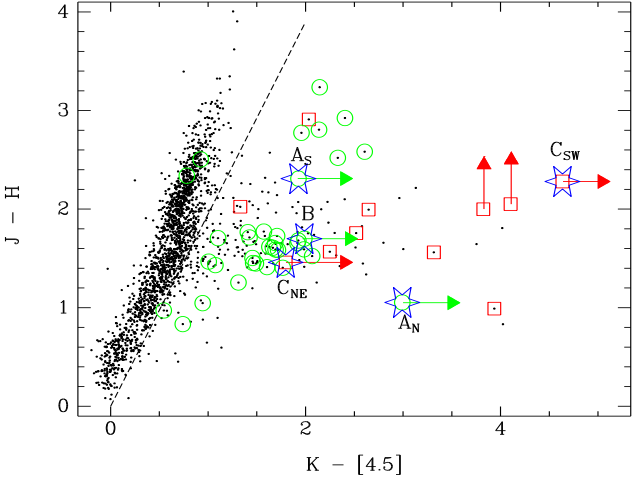}
\caption []{The $J - H/K-[4.5]$ diagram for the sample of 2116 sources with photometric errors
$<$ 0.2 mag in these bands (black dots). The already classified YSO candidates are shown
with red squares (Class~I) and green circles (Class~II). The location of the bright sources
$A$, $B$ and $C$, of which $A$ and $C$ are resolved doubles in the near-IR, are marked with
blue stars.
Some sources have upper limits only in the $J$ band. Some have lower limits due to
saturation in the 4.5 $\mu$m band. The slope of the reddening band is indicated as a dashed
line. }
\label{jhk2}
\end{figure}

\begin{figure}[t]
\includegraphics[angle=270,width=9cm]{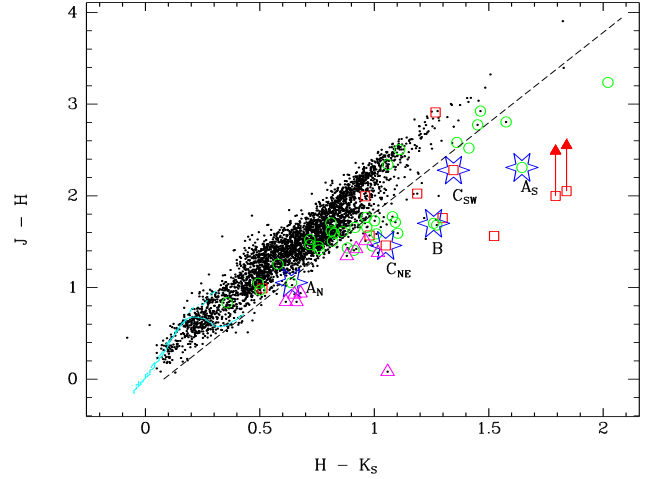}
\caption []{Omega2000 $J-H/H-K_S$ diagram for 3435 sources with errors $<$ 0.06~mag in all bands
(black dots). The empirical reddening vector (dashed line) has a slope of 1.97. We have overlaid
the positions of known Class~Is (red squares), Class~IIs (green circles), and the stars $A$, $B$ 
and $C$ (blue stars). The nine new YSO candidates (magenta triangles) are sources with $K$-band 
excess not already extracted using the previous two CC diagrams.}
\label{jhk}
\end{figure}

Our strategy to define the YSO sample is summarised as follows: 1) find Class~I and Class~II
candidates from the IRAC colour-colour (CC) diagram in Fig.~\ref{irac1234}, 2) use the CC diagram
$J - H/K_S - [4.5]$ in Fig.~\ref{jhk2} to check validity of the candidates found above as well as
extract IR-excess sources not detected in step 1, and 3) use the CC diagram $J - H$/$H - K_S$ in
Fig.~\ref{jhk} to search for any additional IR-excess sources not found in any of the above two
steps.

\begin{figure}
\includegraphics[angle=0,width=9cm]{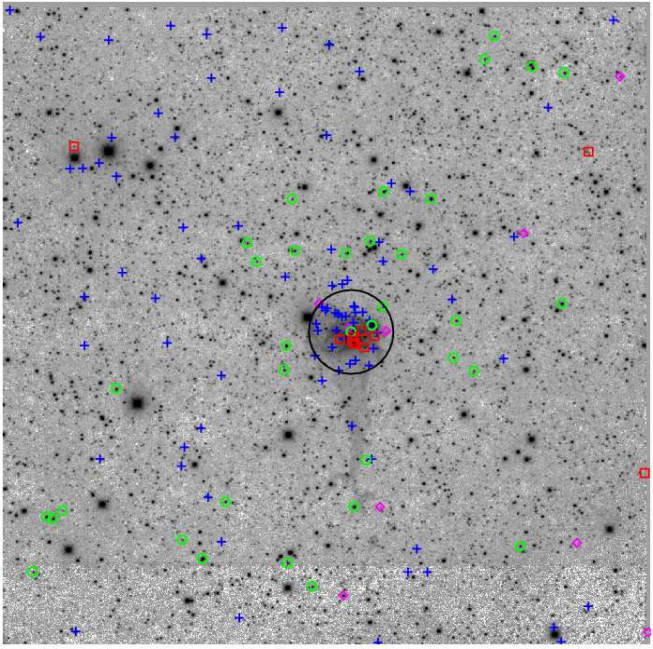}
\caption []{Spatial distribution of the YSO candidates overlaid on the
  Omega2000 15$'\times$15$'$ H$_2$ v=1-0 S(1) image. Class~Is (red
  squares), Class~IIs (green circles), additional YSOs with 4.5 $\mu$m
  excess (blue plus signs) and $K$-band excesses (magenta diamonds)
  not otherwise classified. The black circle has a radius of 1$'$ (0.4
  pc) and shows the approximate extent of the cluster in the globule
  head.}
\label{h2image_ysos}
\end{figure}

From the complete source table with Omega2000 $J$, $H$ and $K_S$ magnitudes and IRAC [3.6], [4.5],
[5.8] and [8.0] magnitudes, we first consider the sample of 302 sources with photometric errors
$<$ 0.2 mag in all IRAC bands, and use the IRAC $[3.6]-[4.5]/[5.8]-[8.0]$ diagram shown in
Fig.~\ref{irac1234} to select Class~I and Class~II candidates. The extinction vector is based
on the average IRAC extinction relations found by \citet{ind05} for the diffuse ISM.
In line with the Class~I and Class~II criteria of \citet{meg04} and \citet{all04} we define
Class~I candidates as those that have:
($[5.8]-[8.0] - \sigma_{[5.8]-[8.0]} >$ 1.2)  and
($[3.6]-[4.5] - \sigma_{[3.6]-[4.5]} >$ 0.4)
or
($[3.6]-[4.5] - \sigma_{[3.6]-[4.5]} >$ 0.8).
This reveals 11 Class~I candidates, of which 8 are located in the globule head within
a radius of 0.4 pc ($\sim$1$'$ at d=1.4 kpc) from the central bright star~A, see Fig.~\ref{h2image_ysos}.
The three remaining sources are faint (13.4 $<$ [3.6] $<$ 14.2 mag) and separated from the globule by 
more than 6$'$. These could be scattered Class~Is anywhere along the line of sight, or they could
be extragalactic sources, such as AGNs or galaxies with strong PAH features contaminating the sample.
We did not perform the rigorous `weeding out' of extragalactic contaminants described in \citet{gut08},
since our main target of this study is a small globule over whose area such a contamination must be
negligible. Some of the Class~I candidates in the globule are very bright in the IRAC bands and
enter the saturation limits. For these, the IRAC colour indices are uncertain, but as shown below, 
their lower limit on the $K - [4.5]$ index strongly supports a Class~I designation.

From the remaining sources we define Class~II candidates as those that satisfy both
$[5.8]-[8.0] - \sigma_{[3.6]-[4.5]} >$ 0.4  and $[3.6]-[4.5] - \sigma_{[3.6]-[4.5]} >$ 0.
This results in 35 Class~II sources of which only two are located in the globule head, namely the
bright stars A and B. Both of these are saturated in all IRAC bands, thus the Class~II designation
is uncertain, but the clear near-IR excesses confirm their YSO status. The remaining
33 Class~IIs have a more extended spatial distribution, as shown in Fig.~\ref{h2image_ysos}.

To check that the above Class~I and Class~II candidates are not subject to PAH contamination
that would masquerade as excess emission at 5.8 and 8.0 $\mu$m, or source confusion in the centre of
the globule, we require that the candidates have point source counterparts in at least two of the
near-IR bands, and we use the $J - H/K - [4.5]$ diagram shown in Fig.~\ref{jhk2} to verify IR-excess
in the $K - [4.5]$ index. Practically all YSO candidates, except three Class~IIs, are well separated
from the reddening band with clear IR-excesses, and in addition, Class~Is have, in general, larger
IR-excesses than Class~IIs. We note that for two Class~Is we only have lower-limit estimates on the
$J$-band magnitudes. For some of the brightest sources, the IRAC magnitudes are upper limits due to
saturation/non-linearity. The double sources Stars A and C are not spatially resolved by IRAC, and we
simply divided their IRAC fluxes in two, thus, their location in this diagram is only an indication.

The $J - H/K - [4.5]$ diagram is also used to find fainter IR-excess sources. It shows a total
sample of 2116 point sources with photometric errors $<$ 0.2 mag in all the bands $J$, $H$, $K$, and
$[4.5]$. The vast majority is located along a reddening band with no excess in the $K-[4.5]$ index.
The upper part of the reddening band becoming bluer is believed to be due to giants with strong CO
absorption in the [4.5] band \citep{meg04}. The reddening vector is based on $A_{\rm 4.5} = 0.43 A_K$
\citep{ind05} together with our near-IR extinction (see below) to have a slope of 1.95 which is in
good agreement with the value 1.97 used by \citet{riv15} as the division line for sources without IR
excesses in a similar study in Cygnus. We assign as IR-excess sources those that are located to the
right and below this slope by more than 2$\sigma$ in the respective colour indices. This gives 130
YSO candidates, of which 88 are new, while 42 of the 44 already suggested YSO candidates also obey
this criterion. The 88 new YSOs have clear IR-excess in their $K - [4.5]$ colour, and the majority
occupy the same locus as previously defined Class~II sources. We interpret these sources as fainter 
Class~II sources that could not be classified with IRAC photometry alone.
As many as 27 of the 88 new YSO candidates are located within a projected radius of $\sim$1$'$ in the
globule head.

Next we explore the sample of 3435 sources with photometric errors $<$ 0.06 mag in $J$, $H$, and $K_S$
to check if there are fainter YSOs with excess emission in the $K_S$ band not already classified above.
The $J-H$/$H-K_S$ diagram is shown in Fig.~\ref{jhk}. The reddening slope was found, by a linear fit to
all sources that show no IR excesses, to be 1.97 $\pm$ 0.01, and it is drawn (dashed line) from the
position of the reddest M dwarfs, that is, shifted by 0.08~mag to the right. This slope corresponds to
$\beta = 1.83$ for the near-IR extinction law in the form $A_{\lambda} \propto \lambda^{-\beta}$.
The positions of known Class~I and Class~II sources are indicated with red squares and green circles,
as well as the locations of the stars $A_N$, $A_S$, $B$, $C_{NE}$ and $C_{SW}$ (blue stars). We identify
sources with $K$-band excess as those lying more than $2\sigma$ of the uncertainty in the colour indices
to the right of and below the reddening vector. A total of 26 such sources were found, of which 16 also
have 4.5 $\mu$m excess and one is a previously found Class~II. Thus, only nine are new YSOs not already
classified.

\begin{figure}[ht]
\includegraphics[angle=0,width=9cm]{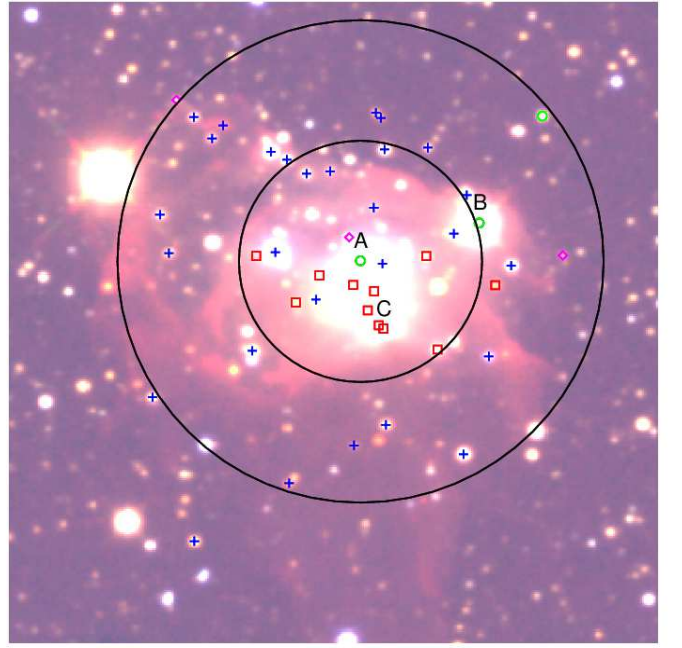}
\caption [] {Close-up view of the globule in RGB-coded colours where red is the $H_2$
image, green is $H$ and blue is $J$ from Omega2000, N up E left. YSOs are marked as in
Fig.~\ref{h2image_ysos} and the three bright stars $A$, $B$ and $C$ are labeled above
(north of) the stars. The two circles indicate radii of 0.2 and 0.4 pc from star $A$.}
\label{h2hj_zoom_ysos}
\end{figure}

In the 15$\arcmin$ $\times$ 15$\arcmin$ field we thus found 11 Class~I candidates,
36 Class~IIs, 88 YSO candidates with excess emission at 4.5 $\mu$m and not already
in the previous two groups, and further 9 sources with K-band excess not previously
designated with a YSO status, making a total of 144 YSO candidates over the whole
field. As shown in Fig.~\ref{h2image_ysos} the spatial density of these is clearly
enhanced at the location of the globule, where we find 40 YSOs within a projected
radius of 1 arc minute (0.4~pc). Fig.~\ref{h2hj_zoom_ysos} shows a close-up view of 
the globule
with the YSOs overlaid and including the three bright stars $A$, $B$ and $C$.
{ Thus, the globule clearly hosts a small embedded
  cluster, supporting the scenario proposed by \citet{sch12}.}

\subsubsection{The YSO sample\label{yso_sample}}

We note that the above Class~I criteria make no distinction between flat-spectrum, Class~I
or Class~0 sources. Those in the Class~I group that are located in the globule, have a
relatively large $[5.8]-[8.0]$ index, combined with a relatively small $[3.6]-[4.5]$ index,
a region in the CC diagram typically occupied by flat-spectrum sources, as shown in
\citet{all07} and \citet{eva09}. A few Class~Is have $K - [4.5]$ indices typical of
Class~II sources.

The sample of 88 + 9 IR-excess sources that can not be given a definite Class~I or Class~II
designation are most likely pre-main sequence (PMS) stars with disks. One of these sources
is 2MASS J20333684+4009387, located approximately 2.5$'$ away from the globule to the WNW, a 
previously spectroscopically identified TTauri star \citep{vin08}. In the $J - H$/$K_S - [4.5]$
diagram, the majority of the IR-excess sources are located in the region dominated by Class~IIs,
while only a few have significantly larger excesses. Thus, we include these in the Class~II group
for statistical considerations. Because our selection of YSO candidates is based on IR-excess only,
a number of sources without IR-excess could, in principle, be young PMS members of Class~III-type. 
The number of sources within 0.4 pc of star $A$ that have detections in all near-IR bands is 140, 
while 40 of these are IR-excess YSOs. For two reference fields of equal size outside the globule, 
we find 86 and 122 sources with detections in all near-IR bands. Thus, on average, there is no 
over-density of stars at the location of the globule, apart from the IR-excess YSOs; an
indication that the contribution of Class~III sources is insignificant.
Keeping in mind the above loose definitions of the Class~I and Class~II groups, we find a number
ratio of Class~I/Class~II in the total area of 11/133, while within a projected radius of
0.4~pc of the globule head this ratio is 8/32. If we consider the inner part of the globule
within a radius of 0.2~pc, this number ratio is 7/10. The lifetime of the Class~I stage is
estimated to be 0.54 Myr \citep{eva09} so that such { a high number of Class~Is suggests that
there has been a recent burst of star formation in the central part of the globule.}

The Class~I sources are all located in the southern part of the inner cluster, while the more evolved
sources are mainly in the northern part. Counting up the total number of YSOs and the Class~I 
fraction, we find that in the northern half of the area inside a radius of 0.4 pc, Class Is constitute 
only 8\% (2/25) of the YSO population, while in the southern half they comprise 53\% (9/17). The 
projected spatial distribution of the Class~Is is highly clustered and clearly to the south of the 
cluster centre. The remaining YSOs, on the other hand, have a projected distribution that is more 
dispersed overall, but 23 of the 31 sources are in the northern part. Despite the fact that the velocity 
dispersion of the newly formed stars will tend to, with time, smear out any initial spatial pattern of 
star formation (up to 1~pc, or $2'5$ on the plane of the sky at the distance of Cygnus~OB2 over 1~Myr 
for a velocity of 1~km~s$^{-1}$), the subsisting predominance of Class~II sources ahead (North) of the 
globule could thus be an indication of star formation progressing from the North towards the south 
inside the globule, that is, propagating away from the center of Cygnus~OB2.

In Table~\ref{tab-ysos} we list the 40 YSOs within a projected radius of 0.4~pc centred on star A
with positions and multi-band photometry.
The near-IR photometry is from Omega2000 except for the brightest sources that enter the
non-linear regime of the detector, where we instead use the NOTCam magnitudes when available
and elsewhere the 2MASS magnitudes. For sources unresolved by Omega2000, we use the NOTCam
photometry where available.

\begin{figure}[ht]
\includegraphics[angle=0,width=8cm]{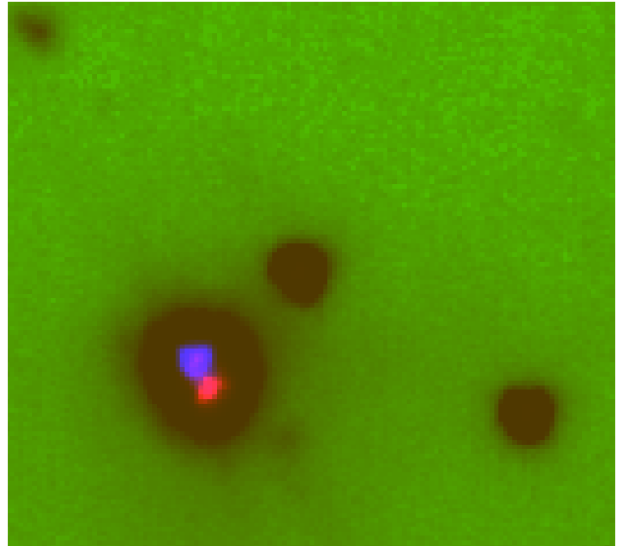}
\caption [] {NOTCam JHK$_S$ 9'' zoom-in on star $A$, resolved in
two components with separation 0$\farcs$52. The southern red component is a NIR
excess source with an extinction A$_K \sim$ 2 mag, while the blue northern
component has no NIR excess and an extinction of $\sim$ 1 mag.}
\label{notcam-zoom}
\end{figure}

\begin{figure}
\includegraphics[angle=270,width=9cm]{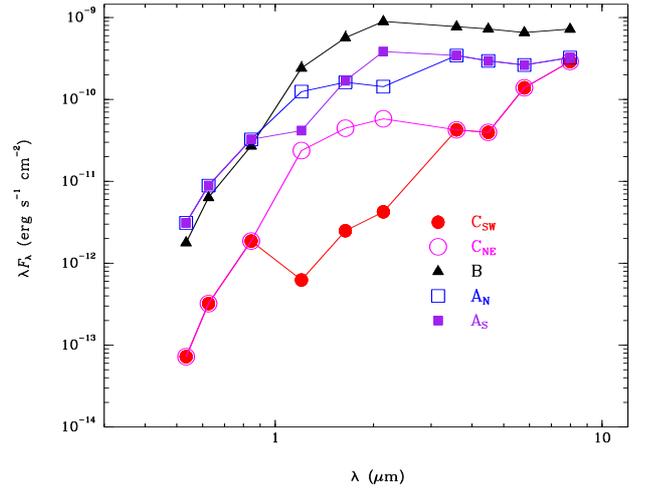}
\caption [] { Observed SEDs of the stars A, B and C. The double
sources A (0$\farcs$52 separation) and C (1$\farcs$42 separation) are resolved only in the bands
$J$, $H$ and $K_S$. The optical counterparts are likely $C_{NE}$ and $A_N$. }
\label{sed}
\end{figure}

\begin{landscape}

{\scriptsize
\begin{table}
\caption{The 40 YSO candidates within a radius of 0.4 pc from star $A$. The columns give:
ID, J2000 positions, magnitudes and errors in all bands: $V$, $R$, $I$, $J$, $H$, $K_S$,
[3.6], [4.5], [5.8], [8.0], and YSO classification. Near-IR photometry is from Omega2000
except for very bright or unresolved sources, where we use the NOTCam photometry. IRAC
magnitudes that reach saturation are flagged as lower limits. The YSOs are listed in groups,
first the 8 Class~Is, then the 4 Class~IIs, followed by 28 PMS stars with IR-excess
emission. The last column gives the YSO class or the selection criterion, that is, excess in
the $K-[4.5]$ index (4) or excess in the $H-K_S$ index (5). The full table of 144 YSO
candidates is available in electronic version.
}
\label{tab-ysos}


\newcommand\cola {\null}
\newcommand\colb {&}
\newcommand\colc {&}
\newcommand\cold {&}
\newcommand\colf {&}
\newcommand\colh {&}
\newcommand\colj {&}
\newcommand\coll {&}
\newcommand\coln {&}
\newcommand\colp {&}
\newcommand\colr {&}
\newcommand\colt {&}
\newcommand\colv {&}
\newcommand\colx {&}
\newcommand\eol{\\}
\newcommand\extline{&&&&&&&&&&&&&\eol}

\begin{tabular}{lrrrrrrrrrrrrll}

\hline \hline
\cola ID\colb $\alpha_{2000}$\colc $\delta_{2000}$\cold $V$\colf $R$\colh $I$\colj $J$\coll $H$\coln $K_S$\colp [3.6]\colr [4.5]\colt [5.8]\colv [8.0]\colx YSO \eol
\cola   \colb hours         \colc deg          \cold mag\colf mag\colh mag\colj mag\coll mag\coln   mag\colp   mag\colr   mag\colt   mag\colv   mag\colx type \eol
\hline
\extline
\cola   3\colb 20:33:46.41\colc 40:08:45.6\cold $>$21.52\colf 20.37$\pm$0.25\colh 20.22$\pm$0.89\colj 19.20$\pm$0.09\coll 16.29$\pm$0.03\coln 15.03$\pm$0.05\colp 13.76$\pm$0.18\colr 12.99$\pm$0.15\colt 10.76$\pm$0.16\colv  8.88$\pm$0.12\colx I\eol
\cola   4\colb 20:33:47.65\colc 40:08:29.5\cold   - \colf   - \colh   - \colj 15.45$\pm$0.05\coll 13.88$\pm$0.03\coln 12.90$\pm$0.04\colp 11.23$\pm$0.06\colr 10.65$\pm$0.05\colt  9.64$\pm$0.09\colv  7.87$\pm$0.13\colx I\eol
\cola   5\colb 20:33:47.90\colc 40:08:52.8\cold   - \colf   - \colh   - \colj 16.66$\pm$0.05\coll 14.64$\pm$0.03\coln 13.45$\pm$0.04\colp 12.69$\pm$0.14\colr 12.12$\pm$0.08\colt  9.69$\pm$0.20\colv  7.78$\pm$0.19\colx I\eol
\cola  $C_{SW}^{ab}$\colb 20:33:48.82\colc 40:08:34.7\cold   - \colf   - \colh   - \colj 16.98$\pm$0.12\coll 14.70$\pm$0.06\coln 13.36$\pm$0.05\colp $<$9.38\colr $<$8.72\colt $<$6.61\colv  $<$4.85\colx I\eol
\cola  $C_{NE}^{ab}$\colb 20:33:48.92\colc 40:08:35.5\cold 21.12$\pm$0.08\colf 19.14$\pm$0.03\colh 16.65$\pm$0.02\colj 13.03$\pm$0.05\coll 11.57$\pm$0.04\coln 10.52$\pm$0.04\colp $<$9.38\colr  $<$8.72\colt  $<$6.61\colv  $<$4.85\colx I\eol
\cola   8\colb 20:33:49.03\colc 40:08:44.0\cold   - \colf   - \colh   - \colj 15.51$\pm$0.09\coll 13.95$\pm$0.07\coln 12.43$\pm$0.08\colp  $<$9.75\colr  $<$9.11\colt  $<$6.43\colv  $<$4.68\colx I\eol
\cola   9\colb 20:33:49.16\colc 40:08:39.2\cold   - \colf   - \colh   - \colj $>$16.47\coll 14.47$\pm$0.11\coln 12.68$\pm$0.07\colp  $<$9.65\colr  $<$8.86\colt  6.70$\pm$0.11\colv  $<$5.06\colx I\eol
\cola  10\colb 20:33:50.71\colc 40:08:41.1\cold   - \colf   - \colh   - \colj 17.26$\pm$0.07\coll 15.50$\pm$0.03\coln 14.20$\pm$0.04\colp 12.35$\pm$0.14\colr 11.68$\pm$0.10\colt  9.12$\pm$0.12\colv  7.18$\pm$0.14\colx I\eol
\cola  24\colb 20:33:45.43\colc 40:09:27.8\cold   - \colf   - \colh   - \colj 16.20$\pm$0.04\coll 14.68$\pm$0.02\coln 13.96$\pm$0.04\colp 12.55$\pm$0.08\colr 11.89$\pm$0.05\colt 11.27$\pm$0.07\colv 10.39$\pm$0.08\colx II\eol
\cola  $B^{a}$\colb 20:33:46.77\colc 40:09:01.1\cold 17.64$\pm$0.01\colf 15.90$\pm$0.01\colh 13.75$\pm$0.01\colj 10.51$\pm$0.04\coll  8.81$\pm$0.02\coln  7.55$\pm$0.04\colp $<$6.23\colr  $<$5.57 \colt  $<$4.93\colv  $<$3.86\colx II\eol
\cola  $A_N^{ab}$\colb 20:33:49.33\colc 40:08:51.7\cold 17.03$\pm$0.01\colf 15.54$\pm$0.01\colh 13.55$\pm$0.01\colj 11.23$\pm$0.08\coll 10.17$\pm$0.06\coln  9.54$\pm$0.06\colp  $<$7.11\colr  $<$6.54\colt  $<$5.92\colv  $<$4.73\colx II\eol
\cola  $A_S^{ab}$\colb 20:33:49.32\colc 40:08:51.3\cold   - \colf   - \colh   - \colj 12.42$\pm$0.08 \coll 10.12$\pm$0.06\coln  8.47$\pm$0.06\colp  $<$7.11\colr   $<$6.54\colt  $<$5.92\colv  $<$4.73\colx II\eol
\cola  65\colb 20:33:46.08\colc 40:08:50.5\cold $>$21.25\colf $>$20.88\colh 18.65$\pm$0.08\colj 15.29$\pm$0.04\coll 13.83$\pm$0.03\coln 13.16$\pm$0.04\colp 11.97$\pm$0.09\colr 11.66$\pm$0.11\colt  9.67$\pm$0.14\colv  7.91$\pm$0.14\colx 4\eol
\cola  66\colb 20:33:46.54\colc 40:08:27.9\cold 20.13$\pm$0.03\colf 19.14$\pm$0.02\colh 17.30$\pm$0.03\colj 15.63$\pm$0.04\coll 15.03$\pm$0.03\coln 14.66$\pm$0.04\colp 13.48$\pm$0.19\colr 13.53$\pm$0.18\colt 10.75$\pm$0.15\colv  8.98$\pm$0.15\colx 4\eol
\cola  68$^{e}$\colb 20:33:47.05\colc 40:09:08.0\cold   - \colf   - \colh   - \colj 15.49$\pm$0.06\coll 14.23$\pm$0.05\coln 13.57$\pm$0.05\colp 11.76$\pm$0.11\colr 11.46$\pm$0.11\colt  9.41$\pm$0.13\colv  7.25$\pm$0.17\colx 4\eol
\cola  69\colb 20:33:47.08\colc 40:08:03.5\cold $>$22.49\colf 21.24$\pm$0.08\colh 18.73$\pm$0.05\colj 15.19$\pm$0.04\coll 13.46$\pm$0.03\coln 12.51$\pm$0.04\colp 11.37$\pm$0.05\colr 10.93$\pm$0.05\colt 10.76$\pm$0.14\colv 10.49$\pm$0.23\colx 4\eol
\cola  70\colb 20:33:47.31\colc 40:08:58.4\cold   - \colf   - \colh   - \colj 16.44$\pm$0.06\coll 14.34$\pm$0.05\coln 13.10$\pm$0.04\colp 11.17$\pm$0.10\colr 10.70$\pm$0.06\colt  6.54$\pm$0.04\colv  7.43$\pm$0.12\colx 4\eol
\cola  71\colb 20:33:47.88\colc 40:09:19.7\cold   - \colf   - \colh   - \colj 17.06$\pm$0.04\coll 15.45$\pm$0.03\coln 14.58$\pm$0.04\colp 13.81$\pm$0.17\colr 13.27$\pm$0.16\colt   - \colv   - \colx 4\eol
\cola  73\colb 20:33:48.75\colc 40:08:10.7\cold   - \colf   - \colh   - \colj 16.83$\pm$0.04\coll 14.88$\pm$0.03\coln 13.74$\pm$0.04\colp 12.31$\pm$0.07\colr 11.83$\pm$0.07\colt 11.48$\pm$0.14\colv  \colx 4\eol
\cola  74\colb 20:33:48.82\colc 40:09:19.2\cold   - \colf   - \colh   - \colj 15.63$\pm$0.04\coll 14.15$\pm$0.02\coln 13.36$\pm$0.04\colp 11.89$\pm$0.11\colr 11.42$\pm$0.07\colt   - \colv   - \colx 4\eol
\cola  75\colb 20:33:48.85\colc 40:08:50.8\cold 21.31$\pm$0.09\colf 19.50$\pm$0.03\colh 17.21$\pm$0.03\colj 13.79$\pm$0.05\coll 12.33$\pm$0.04\coln 11.49$\pm$0.04\colp 10.31$\pm$0.07\colr  9.83$\pm$0.05\colt  7.61$\pm$0.12\colv  $<$5.57\colx 4\eol
\cola  76\colb 20:33:48.89\colc 40:09:27.1\cold   - \colf   - \colh   - \colj 18.80$\pm$0.06\coll 16.80$\pm$0.03\coln 15.60$\pm$0.04\colp 14.02$\pm$0.19\colr 13.44$\pm$0.14\colt   - \colv   - \colx 4\eol
\cola  77\colb 20:33:49.00\colc 40:09:28.3\cold   - \colf   - \colh   - \colj 18.22$\pm$0.05\coll 16.07$\pm$0.03\coln 15.15$\pm$0.04\colp 14.02$\pm$0.19\colr 13.44$\pm$0.14\colt   - \colv   - \colx 4\eol
\cola  78\colb 20:33:49.04\colc 40:09:04.8\cold   - \colf   - \colh   - \colj 17.41$\pm$0.06\coll 15.97$\pm$0.03\coln 14.96$\pm$0.04\colp 12.72$\pm$0.16\colr 12.37$\pm$0.19\colt  9.70$\pm$0.13\colv  8.18$\pm$0.10\colx 4\eol
\cola  80\colb 20:33:49.44\colc 40:08:05.6\cold   - \colf   - \colh   - \colj 18.52$\pm$0.05\coll 16.86$\pm$0.03\coln 15.83$\pm$0.04\colp 14.11$\pm$0.17\colr 13.03$\pm$0.13\colt 13.25$\pm$0.67\colv   - \colx 4\eol
\cola  82\colb 20:33:49.98\colc 40:09:13.7\cold   - \colf   - \colh   - \colj 17.35$\pm$0.05\coll 15.94$\pm$0.03\coln 15.14$\pm$0.05\colp 13.26$\pm$0.20\colr 13.16$\pm$0.19\colt 10.63$\pm$0.15\colv  8.93$\pm$0.15\colx 4\eol
\cola  83\colb 20:33:50.27\colc 40:08:41.8\cold   - \colf   - \colh   - \colj 16.89$\pm$0.08\coll 14.82$\pm$0.03\coln 13.32$\pm$0.04\colp 11.65$\pm$0.14\colr 11.09$\pm$0.18\colt  8.04$\pm$0.13\colv  $<$6.34\colx 4\eol
\cola  85\colb 20:33:50.49\colc 40:09:13.1\cold   - \colf   - \colh   - \colj 16.49$\pm$0.04\coll 14.73$\pm$0.03\coln 13.69$\pm$0.04\colp 12.26$\pm$0.10\colr 11.77$\pm$0.08\colt 10.63$\pm$0.12\colv  9.04$\pm$0.13\colx 4\eol
\cola  86\colb 20:33:50.83\colc 40:07:56.1\cold   - \colf   - \colh   - \colj 18.65$\pm$0.06\coll 16.38$\pm$0.03\coln 15.19$\pm$0.04\colp 13.84$\pm$0.17\colr 13.50$\pm$0.18\colt 12.86$\pm$0.35\colv   - \colx 4\eol
\cola  87\colb 20:33:50.92\colc 40:09:16.6\cold   - \colf   - \colh   - \colj 15.88$\pm$0.04\coll 14.23$\pm$0.03\coln 13.29$\pm$0.04\colp 11.99$\pm$0.09\colr 11.54$\pm$0.06\colt  9.85$\pm$0.09\colv  8.25$\pm$0.12\colx 4\eol
\cola  88\colb 20:33:51.15\colc 40:08:53.5\cold 18.35$\pm$0.02\colf 16.72$\pm$0.01\colh 14.62$\pm$0.01\colj 12.15$\pm$0.04\coll 11.27$\pm$0.02\coln 10.77$\pm$0.04\colp 10.28$\pm$0.05\colr  9.82$\pm$0.06\colt  8.62$\pm$0.11\colv  7.00$\pm$0.10\colx 4\eol
\cola  89\colb 20:33:51.26\colc 40:09:18.5\cold $>$21.82\colf $>$20.31\colh 18.66$\pm$0.07\colj 14.76$\pm$0.04\coll 13.07$\pm$0.03\coln 12.29$\pm$0.04\colp 11.34$\pm$0.09\colr 11.13$\pm$0.04\colt  9.26$\pm$0.09\colv  7.55$\pm$0.14\colx 4\eol
\cola  90\colb 20:33:51.63\colc 40:08:29.0\cold 19.71$\pm$0.03\colf 18.65$\pm$0.02\colh 17.27$\pm$0.02\colj 15.79$\pm$0.04\coll 13.75$\pm$0.03\coln 12.71$\pm$0.04\colp 11.75$\pm$0.05\colr 11.41$\pm$0.08\colt 11.30$\pm$0.20\colv   - \colx 4\eol
\cola  94\colb 20:33:52.29\colc 40:09:24.9\cold   \colf   - \colh   - \colj 19.19$\pm$0.07\coll 17.08$\pm$0.03\coln 15.72$\pm$0.04\colp 12.77$\pm$0.14\colr 12.76$\pm$0.12\colt  9.99$\pm$0.11 \colv  8.23$\pm$0.10\colx 4\eol
\cola  96\colb 20:33:52.53\colc 40:09:21.7\cold $>$22.00\colf $>$21.69\colh 20.47$\pm$0.16\colj 17.05$\pm$0.05\coll 15.51$\pm$0.03\coln 14.67$\pm$0.04\colp 12.96$\pm$0.17\colr 13.08$\pm$0.14\colt 10.09$\pm$0.11\colv  8.37$\pm$0.13\colx 4\eol
\cola  98\colb 20:33:52.92\colc 40:09:27.0\cold   - \colf   - \colh   - \colj 16.64$\pm$0.04\coll 15.05$\pm$0.03\coln 14.26$\pm$0.04\colp 12.96$\pm$0.15\colr 12.91$\pm$0.12\colt 10.06$\pm$0.11\colv  8.38$\pm$0.10\colx 4\eol
\cola  99\colb 20:33:53.45\colc 40:08:53.1\cold   - \colf   - \colh   - \colj 17.25$\pm$0.05\coll 15.37$\pm$0.04\coln 14.38$\pm$0.06\colp 13.32$\pm$0.15\colr 13.05$\pm$0.12\colt 11.42$\pm$0.20\colv  9.60$\pm$0.18\colx 4\eol
\cola 100\colb 20:33:53.64\colc 40:09:02.7\cold $>$21.63\colf $>$21.36\colh 18.73$\pm$0.21\colj 16.41$\pm$0.04\coll 14.88$\pm$0.03\coln 14.13$\pm$0.04\colp 12.72$\pm$0.12\colr 12.85$\pm$0.14\colt 10.25$\pm$0.14\colv  8.41$\pm$0.13\colx 4\eol
\cola 140\colb 20:33:44.97\colc 40:08:53.1\cold   - \colf   - \colh   - \colj 18.26$\pm$0.05\coll 16.84$\pm$0.03\coln 15.91$\pm$0.04\colp   - \colr   - \colt   - \colv   - \colx 5\eol
\cola 142\colb 20:33:49.57\colc 40:08:57.3\cold   - \colf   - \colh   - \colj 16.39$\pm$0.05\coll 15.01$\pm$0.03\coln 13.99$\pm$0.04\colp   - \colr   - \colt   - \colv   - \colx 5\eol
\hline
\end{tabular}

$^{a}$ NOTCam photometry for JHKs bands,
$^{b}$ IRAC unresolved double, flux divided,
$^{e}$ Extended IRAC source.


\end{table}
}
\end{landscape}

\subsection{The brightest members of the globule population \label{ind_stars}}

The three brightest stars projected on the globule at near- and mid-infrared wavelengths are 
respectively labeled as stars~$A$, $B$ and $C$ in Fig.~\ref{h2hj_zoom_ysos}. The spectroscopic 
study of these three stars yields additional information useful to assess their nature as well 
as the evolutionary status of the globule and the stellar population associated with it.

With the high spatial resolution of our NOTCam images, Star~$A$ is resolved into two close 
components separated by 0$\farcs$52, see Fig.~\ref{notcam-zoom}. This is too close to be resolved 
by {\it Spitzer}, and the combined luminosity of the pair results in the characteristics of a 
Class~II source. The resolution of the pair in the NOTCam images at $J$, $H$ and $K_S$ shows that 
only the Southern component, which is also redder, displays $K-$band excess (see Fig.~\ref{jhk}).

Star~$B$ also has a Class~II SED, and appears unresolved in all the images. The extended emission 
of the globule hints at a peak coincident with its position.  The brightest Class~I source in the 
mid-IR is star~$C$, approximately 15$\arcsec$ to the South of star~$A$. Star~$C$ is a resolved 
binary with a separation of 1$\farcs$42 both in the Omega2000 and NOTCam datasets. Both components 
have near-IR excess. The NE component is the brightest and the SW component is the reddest. 
Figure~\ref{sed} shows the observed SEDs of these stars from the $V$ band to the 8 $\mu$m IRAC band.

\begin{figure}
\includegraphics[angle=0,width=9cm]{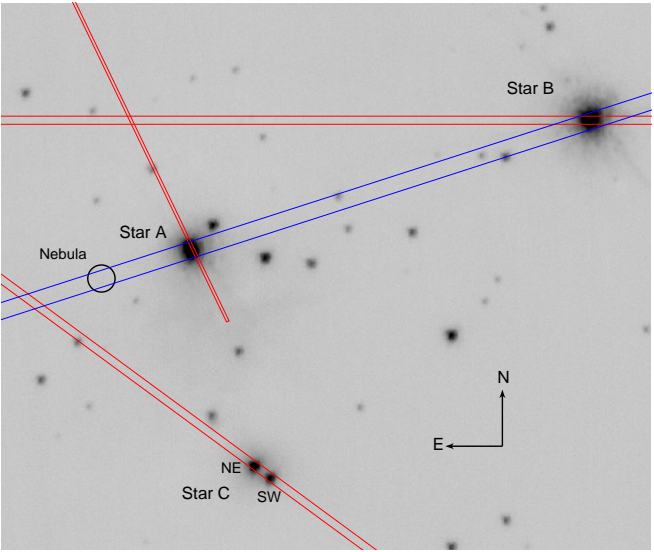}
\caption [] {Slit positions and widths overlaid on NOTCam high-resolution K-band image zoomed
in to 50$\arcsec \times$40$\arcsec$. Three near-IR slit positions (red) across the double
sources stars $A$ and $C$, and the single source $B$, and the optical slit (blue) across
stars~$A$ and $B$ and the Nebula. }
\label{slitimage}
\end{figure}

\begin{figure*}[ht]
\includegraphics[angle=0]{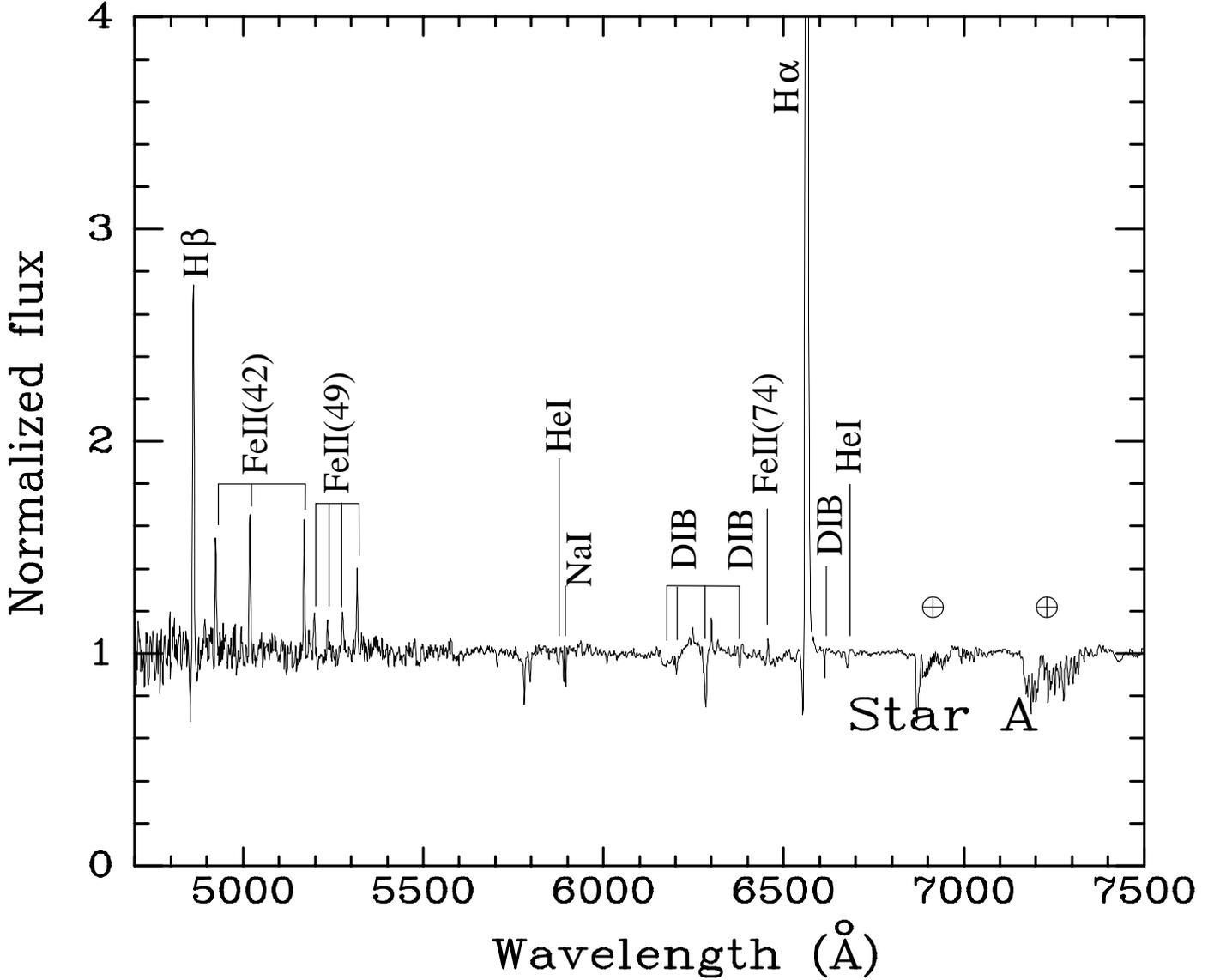}
\caption [] {Optical spectrum of star $A$. Note the absorption lines of Helium at 5875~\AA \ and 6678~\AA .}
\label{starA}
\end{figure*}

\subsubsection{Stars $A$ and $B$}

\begin{figure*}[t]
\includegraphics[angle=0]{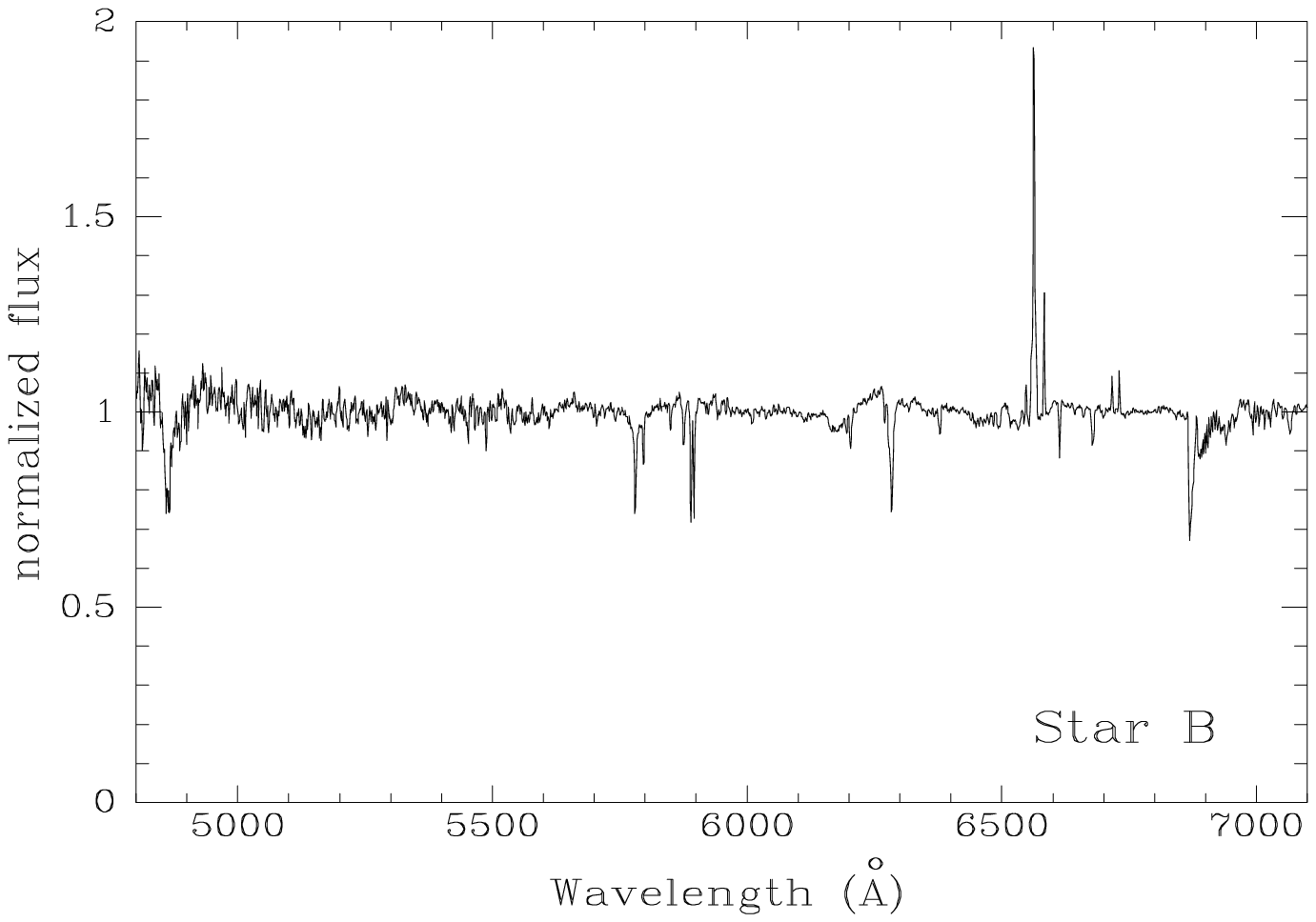}
\caption [] {Optical spectrum of star $B$. Note the absorption lines of Helium at 5875~\AA \ and 6678~\AA .}
\label{starB}
\end{figure*}

The spectra of both stars $A$ and $B$ (see Fig.~\ref{slitimage} for the slit positions and 
Figs.~\ref{starA} and \ref{starB} for the spectra) display very reddened continua consistent with 
the high extinction derived in their direction. Besides the telluric absorption bands in the red, 
both spectra display abundant diffuse interstellar bands \citep{sar06}. However, the 
He\,{\scriptsize I} 5875~\AA \ and 6678~\AA \ lines are clearly seen in absorption in both spectra, 
thus unambiguously classifying {\sl both stars as B-type or earlier}.

The emission-line spectra of both stars show remarkable differences, though. Superimposed onto star 
$B$ is the spectrum of a low-excitation \HII\ region with an electron density $n_e \simeq 800$~cm$^{-3}$ 
derived from the ratio of the [S\,{\scriptsize II}] lines at 6717~\AA \ and 6731~\AA , representing 
a very local enhancement of the \HII\ nebula. The expected photospheric H$\alpha$ absorption in the 
spectrum of star $B$ is completely filled by the nebular emission at the same wavelength, preventing 
the determination of its spectral type through the ratio of the He\,{\scriptsize I} to the Balmer lines. 
However, the ability of star $B$ to excite a local \HII\ region and the low excitation suggests an 
early B spectral type. The possible appearance of weak Si\,{\scriptsize III} absorption at 5740~\AA , 
which is only tentatively detected in our spectrum, would constrain it to the B0.5 - B1.5 interval 
\citep{wal80}.

While the continuum and absorption line spectrum of star~$A$ is similar to that of star~$B$, including 
the presence of He\,{\scriptsize I} in absorption, the emission-line spectrum at its position is 
markedly different and clearly indicates a circumstellar origin. The very intense H$\alpha$ emission 
displays an obvious P-Cygni profile, with an extended red wing. The same P~Cygni profile is clearly 
seen in the H$\beta$ line as well. No other prominent emission lines appear in the red, but the blue 
spectrum displays strong Fe\,{\scriptsize II} lines of multiplets 42 and 49.

The concurrent appearance of H$\alpha$ with P~Cygni profile \citep{bes99,kra08} and the 
Fe\,{\scriptsize II} multiplets in emission \citep{ros99,her04} together with a photospheric spectrum 
displaying He\,{\scriptsize I} in absorption classify {\sl Star~A as a Herbig Be star}. However, the 
optical spectrum needs to be interpreted with caution in view of its binarity. The orientation of the 
slit, coupled with its width, the 1$''$.5 seeing under which the spectra were taken, and the close 
separation between both components implies that the spectrum is the blend of the individual spectra of 
components North and South. The photometry presented in Table~\ref{tab-ysos} suggests that the Northern 
component, being the bluest, dominates the continuum in the region covered by our spectra, but the 
characteristics of component South, being a source with strong near-infrared excess, make it a 
candidate contributor to the emission-line spectrum.

\begin{figure}[t]
\includegraphics[angle=270,width=9cm]{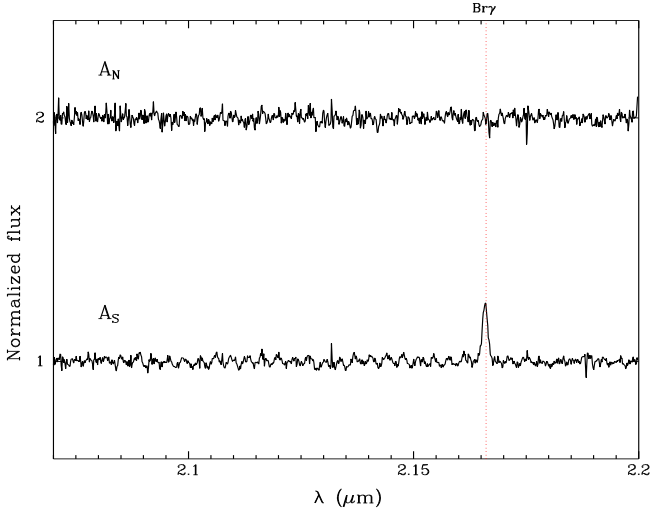}
\caption [] {Medium-resolution (R=5500) normalised $K$-band spectra of the two 0$\farcs$52
separation components of star $A$ resolved spatially in $A_N$ (upper) and $A_S$ (lower). The
spectra are offset by unity on the y-axis.}
\label{kspec2}
\end{figure}

$K$-band spectroscopy of star~$A$ under excellent image quality conditions, allowing us to partly 
resolve the spectra of both components, is presented in Figure~\ref{kspec2}. Both spectra have 
essentially featureless continua, consistent with them being produced by early-type photospheres with
some degree of continuum veiling. The spectra also show that Br$\gamma$ is largely produced by the 
Southern component, with an equivalent width of 8 \AA . The much stronger Br$\gamma$ emission of the 
southern component hints that the H$\alpha$ emission in the blended spectrum, and possibly the entire
emission-line spectrum of the system, is actually dominated by the redder Southern component.

\subsubsection{Star C \label{StarC}}

\begin{figure}[t]
\includegraphics[angle=270,width=9cm]{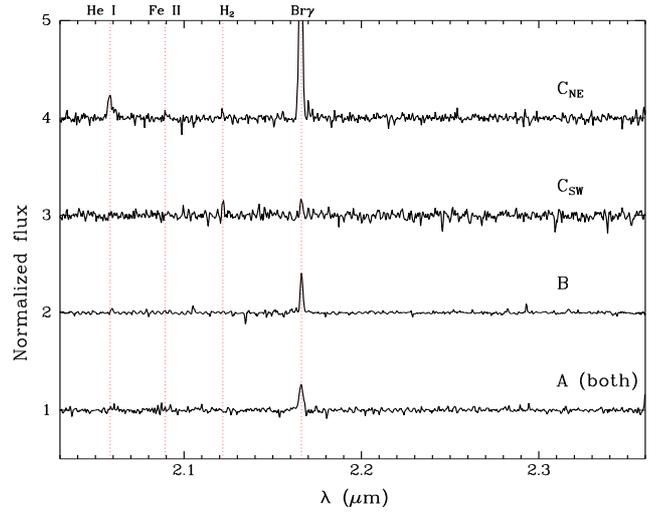}
\caption [] {Low-resolution (R=2100) $K$-band spectra of the three brightest stars. Star~$A$
shows both components unresolved, star~$B$ is apparently single, and star~$C$ is well
resolved into the NE and SW components. The spectra are offset by unity on the y-axis.}
\label{kspec}
\end{figure}

Star~$C$ is one of the brightest stars in the globule at near- and mid-IR wavelengths, with
colours corresponding to a Class~I young stellar object based on {\it Spitzer} photometry, which does
not resolve the two 1$\farcs$4 separation sources. Both components $C_{NE}$ and $C_{SW}$ have
near-IR excesses (cf. Fig.~\ref{jhk})with $C_{SW}$ being the fainter and the redder of the two.
The K-band spectra of the two components of Star $C$ are shown in Fig.~\ref{kspec}.

The brighter northeastern component has a strong Br$\gamma$ emission line with an equivalent
width as large as 70 \AA \ , typically seen only for the most luminous Class\,I type YSOs 
\citep{ish01}, together with He\,{\scriptsize I} at 2.0583 $\mu$m in emission and a tentative 
detection of Fe\,{\scriptsize II} in emission at 2.0893 $\mu$m. The lack of photospheric features 
hampers its classification, but the simultaneous presence of IR excess and He\,{\scriptsize I} and 
Fe\,{\scriptsize II} in emission suggests a massive YSO. The fainter and redder
southwestern component has a faint Br$\gamma$ emission line (EW $\sim$ 3 \AA ) on an otherwise
featureless spectrum. Both stars show faint and slightly extended H$_2$ emission at 2.1218 $\mu$m.
We measure the line ratio of He\,{\scriptsize I} (2.058 $\mu$m) to Br$\gamma$ to be $\sim$ 0.1. 
According to models by \citet{lum03} in their study of He and H line ratios of compact H~II regions, 
a line ratio of 0.1 suggests that $C_{NE}$ has a temperature T$_{\rm eff} \ll $ 36000~K, and probably 
$<$ 30000~K. Thus, $C_{NE}$ could be a late O or early B star, probably more massive than Star~$A$, 
where we do not detect He\,{\scriptsize I} emission. For further comparison, we obtained a K-band 
spectrum of Star~$B$ and also found a He\,{\scriptsize I} to Br$\gamma$ ratio of approximately 0.1 
for this star, but with the difference that the Br$\gamma$ equivalent width of star~$B$ is almost 
a factor of ten smaller (see Fig.~\ref{kspec}).
Interestingly, the He\,{\scriptsize I} emission is seen superimposed on the spectrum of $C_{NE}$ 
alone and is not seen towards $C_{SW}$ located only 1$\farcs$4 ($<$ 0.01~pc projected distance) away.

\subsection{Intrinsic properties of the YSO population\label{phys_ysos}}

The combination of near-infrared photometry and the visible photometry available for many of the
sources in the globule allows us to probe the photosphere-dominated region of their spectral energy
distributions and compare it to evolutionary tracks to infer their intrinsic properties. The number of
YSOs identified in the concentration toward the head of the globule also provides some insight into
their collective properties, most notably their mass function.

We have used the PMS evolutionary tracks of \citet{pal99} for masses below 4~M$_\odot$, with colours 
from \citet{tes98}, complemented for higher masses with the evolutionary tracks of \citet{lej01} for 
non-rotating stars of solar metallicity, which include a synthetic computation of the intrinsic 
colours. The fluxes at $J$ and shorter wavelength bands were fitted to the synthetic fluxes leaving 
the $V$-band extinction $A_V$ as an adjustable parameter, scaling it to the other bands by means
of the Cardelli et al. (1989) extinction law for a total-to-selective extinction ratio $R_V = 3.1$. 
We did not include infrared bands longward from $J$ in the fit due to the possible existence of 
circumstellar emission substantially contributing to the flux at long wavelengths. We used the 1~Myr 
isochrone as being approximately representative of young Class~II objects \citep[e.g.][]{eva09}, 
where the initial circumstellar envelope has been largely dissipated and the central object has become 
accessible to visible and near-infrared observations. Isochrone fitting, as a way to determine the 
age of the aggregate, is not applicable to our data due to the degeneracy of isochrones at high and 
intermediate masses in colour-magnitude diagrams, and to the fact that we do not have optical 
photometry for the lowest-mass stars, for which such degeneracy breaks, which would allow us to 
independently fit the spectral energy distribution and the amount of foreground extinction as 
described below.

  A least-square fit to the $V, R_C, I_C, J$ magnitudes was performed
  for those stars having measurements in at least two of those bands
  to derive their mass and the extinction in their direction. This is
  the case for 13 of the 40 YSO candidates listed in
  Table~\ref{tab-ysos}. Those stars show $A_V$ values mostly within
  the range $7.0< A_V < 11.9$, with the exception of two sources
  without strong infrared excess for which we derive a much smaller
  value of $A_V$, and which are likely foreground stars. The small
  size and low density of the head of the globule suggest a low level
  of internal extinction: assuming a maximum extension along the line
  of sight of 0.8~pc, a typical density $n(H) \sim 200$~cm$^{-3}$ (see
  Section~\ref{neb_spect}) and a proportionality between visible
  extinction $A_V$ and hydrogen column density $N_H$ of the form $A_V
  = 5.6 \times 10^{-22}$~cm$^{2} N_H$ \citep{lis14}, we estimate that
  the internal extinction does not exceed $A_V \simeq 0.3$~mag. Most
  of the extinction is thus produced in the foreground, possibly
  associated with the Cygnus Rift. This feature is an extended
  area of gas and dust at a distance of approximately 600 pc, not related to the
  Cygnus X region \citep[for example]{sch07}, which produces a sudden increase 
  in visual extinction \citep{staude1982}.  For
  stars with no photometric measurements in bands bluer than $J$ we
  have adopted $A_V =9.6$, the average value obtained for those stars
  for which individual $A_V$ can be obtained, and have estimated their
  mass from the corresponding absolute magnitude $M_J$ given by the
  isochrones. The adopted value has an estimated uncertainty of $\sim
  \pm 3$~mag, translating to an uncertainty $M_J \sim \pm 0.8$~mag.

  \begin{table}[t]
\caption{Estimated properties of the YSOs, assuming 1~Myr age.}
\begin{tabular}{lcccl}
\hline \hline
Source & Mass (M$_\odot$) & $M_J$ & $A_V$ & Notes \\
\hline
    3  &  $<0.1$  &   -    & - & \\
    4  &  1.8  &  1.89 &  9.6$^a$  & \\
    5  &  0.7  &  3.10 &  9.6$^a$  & \\
 $C_{NE}$ & 8.1  & -1.12 & 11.9  &  \\
    8  & 1.7  &  1.95 &  9.6$^a$  & \\
    9  & 0.8  &  2.91 &  9.6$^a$  & \\
   10  & 0.4  &  3.70 &  9.6$^a$  & \\
   $B$ & 22.9 & -3.32 & 10.7  &     \\
   $A_N$ & 12.9 & -2.05 &  8.8  &  \\
   65  & 1.3  &  2.28 &  7.8  & \\
   66  & 0.3  &  4.37 &  1.6  & Likely foreground \\
   68  & 1.8  &  1.93 &  9.6$^a$  & \\
   69  & 1.9  &  1.91 &  8.8 & \\
   70  & 0.9  &  2.87 &  9.6$^a$  & \\
   71  & 0.5  &  3.50 &  9.6$^a$  & \\
   73  & 0.6  &  3.27 &  9.6$^a$  & \\
   74  & 1.6  &  2.07 &  9.6$^a$  & \\
   75  & 5.2  & -0.13 & 11.0  & \\
   76  & 0.1  &  5.23 &  9.6$^a$  & \\
   77  & 0.2  &  4.65 &  9.6$^a$  & \\
   78  & 0.4  &  3.85 &  9.6$^a$  & \\
   80  & 0.1  &  4.95 &  9.6$^a$  & \\
   82  & 0.4  &  3.79 &  9.6$^a$  & \\
   83  & 0.6  &  3.33 &  9.6$^a$  & \\
   85  & 0.8  &  2.93 &  9.6$^a$  & \\
   87  & 1.2  &  2.32 &  9.6$^a$  & \\
   88  & 8.9  & -1.21 &  9.1  &  \\
   89  & 2.6  &  0.88 & 10.9  &  \\
   90  &  -   &  -    & $\sim 0$ & Foreground \\
   94  & $<0.1$   & - & - & \\
   96  &  0.3  &  4.28 &  7.0  & \\
   98  &  0.8  &  3.08 &  9.6$^a$  & \\
   99  &  0.4  &  3.69 &  9.6$^a$  & \\
  100  &  0.2  &  4.86 &  2.7  & Foreground? \\
  140  &  0.2  &  4.69 &  9.6$^a$  & \\
  142  &  0.9  &  2.82 &  9.6$^a$  & \\
\hline
\end{tabular}
\label{tab-physprop}

$^a$ Properties based on the $J$ magnitude adopting average value $A_V$ (see text).
\end{table}

The derived mass $M$ and absolute $J$ magnitude $M_J$, and the derived or adopted extinction $A_V$, are
given in Table~\ref{tab-physprop}. The YSO population of the globule covers over two orders of magnitude
in mass, from Stars~$A_N$ and $B$, which are the most massive at $M \simeq 13$~M$_\odot$ and $M \simeq
23$~M$_\odot$, respectively, (in consistency with their early B spectral types) to below 0.1~M$_\odot$, 
as wederive for stars \#3 and \#94. These latter two objects, which display clear infrared excesses at 
the longer {\it Spitzer} bands, may be massive brown dwarf YSOs of the aggregate in the globule.

We note that the mass that we derive for star~$C_{NE}$ of 8.1~M$_\odot$ makes it significantly less 
massive than stars $A_N$ and $B$, in consistency with its lower brightness in the visible and 
near-infrared.
However, such relatively moderate mass for a young stellar object seems to be in contradiction with the
appearance of He\,{\scriptsize I} in emission in the $K$-band spectrum, which we used in 
Sect.~\ref{StarC} to argue for a spectral type possibly earlier than those of stars $A_N$ and $B$. The 
apparent contradiction might be accounted for by the possibility that the He\,I emission is not produced 
in a compact \hii \ region around star~$C$ but in the hot spots caused by accretion onto its surface 
instead, thus being unrelated to the temperature of the central object. On the other hand it is also 
possible that the main contribution to the
flux received at short wavelengths does not actually come directly from the central object, but rather
from the scattering of its light by the dusty circumstellar envelope that is also responsible for the
strong infrared excess of star~$C$ at long wavelengths. The available observations do not allow us to
discern between those two possibilities, and we must consider that the actual mass of star~$C$ remains
particularly uncertain.

\begin{figure}
\includegraphics[angle=0,width=9cm]{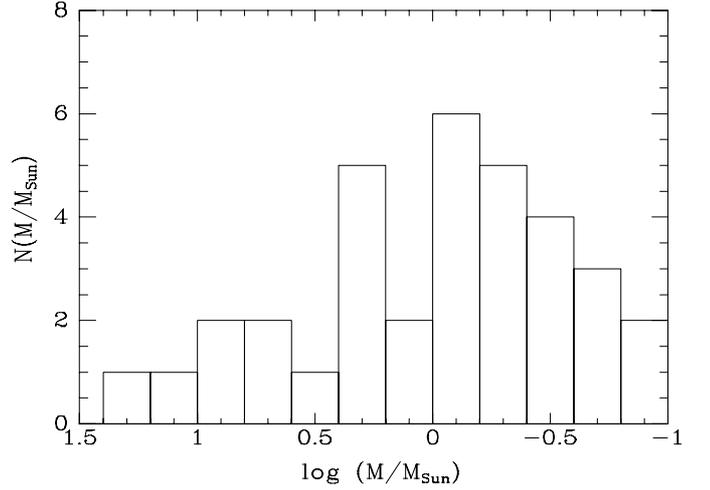}
\caption []{Histogram of the distribution of stellar masses in the globule aggregate, according to the
derived values listed in Table~\ref{tab-physprop}. An age of 1~Myr has been assumed.}
\label{imf}
\end{figure}

The determination of the mass function of the aggregate is complicated by several factors that include
the uncertainties in PMS evolutionary tracks, the unknown age of the members, the approximate
correction by extinction effects, incompleteness at the lowest masses, an incompleteness due to the
existence of members still in early evolutionary phases and not yet accessible to direct observation in
the visible and near-infrared. Nevertheless, the histogram of masses shown in Fig.~\ref{imf} that we
obtain from the estimated physical properties gives some interesting insights into the collective 
properties of the aggregate.

Assuming a power-law slope of the initial mass function of the form $N(M) \propto M^{(\Gamma + 1)} \Delta
\log M$, the histogram for masses above solar indicates a slope $\Gamma = -1.6 \pm 0.2$, shallower than
the Salpeter exponent ($\Gamma = - 2.35$) characterising the functional forms of the initial mass function
above one solar mass. The peak of the mass histogram is, in principle, reminiscent of that observed in the
initial mass function in the field and in clusters, reproduced by the log-normal part of the analytical
representation of the initial mass function \citep{cha05} where such a peak appears at 0.1-0.2~M$_\odot$.
We find it at a significantly higher mass, $\sim 1$~M$_\odot$, but the mass of the peak is very sensitive
to the adopted age and drops to $\sim 0.3$~M$_\odot$ in the extreme case of adopting the 0.1~Myr PMS 
isochrone of \citet{pal99}. The offset may also be a simple artifact of incompleteness of our 
sample, but we note that stars near the peak have relatively bright absolute magnitudes, 
$M_J \simeq 2.5$), well above our completeness limit even at the distance modulus of Cygnus OB2 when 
obscured by $A_V = 10$ ($A_J = 2.9$), therefore suggesting that the peak is real. If our sample is nearly 
complete down to where the initial mass function declines with decreasing mass, the sum of the estimated 
stellar masses given in Table~\ref{tab-physprop} yields a total mass of the stellar aggregate of $\sim 
90$~M$_\odot$.

An overestimate of the actual age of the aggregate members would also have the effect of making the 
high-mass slope of the mass function shallower: using the 0.1~Myr isochrone yields $\Gamma = -1.4 
\pm 0.2$. 
All this hints to a departure from the canonical initial mass function for the aggregate, in the sense
that the mass function of the population embedded in IRAS~20319+3958 would be top-heavy and have its 
peak at a higher mass. However, given the many uncertainties involved, deeper imaging 
and complementary observations allowing a more robust determination of the mass distribution would be 
required to make firm conclusions on this point.

\subsection{The extended emission}

\begin{figure}[ht]
\includegraphics[angle=0,width=9cm]{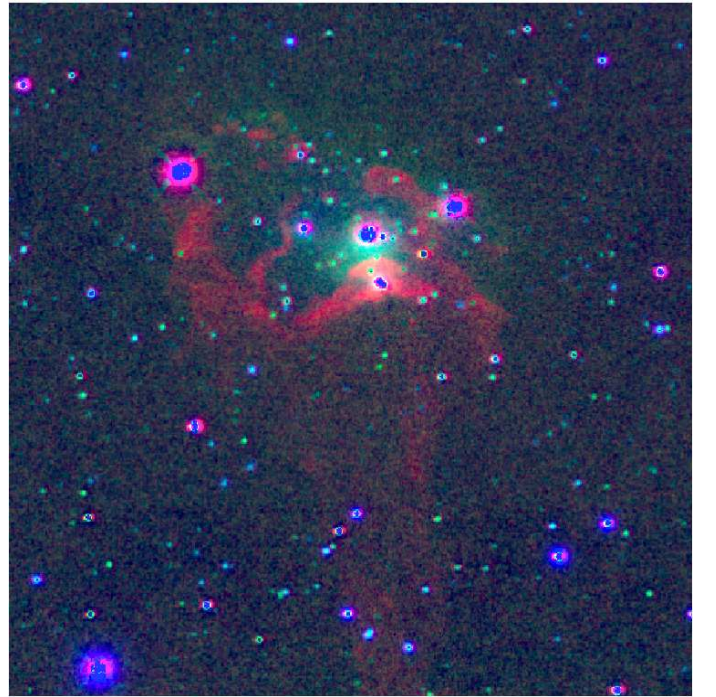}
\caption [] {The globule colour-coded with H$_2$ emission (red), Br$\gamma$ (green), and J-band 
(blue).}
\label{nb_rgb_im}
\end{figure}

Deep narrow-band imaging at 2.122~$\mu$m, covering the ro-vibrational 1-0 S(1) line of H$_2$, and at
2.166~$\mu$m, covering Br$\gamma$, shows extended emission in both of these lines at the location of
the globule (see Fig.~\ref{nb_rgb_im}). We have made approximate continuum subtraction by scaling the
K$_S$ band image intensity down by a factor 0.075 and 0.055 for H$_2$ and Br$\gamma$, respectively, 
based on the transmission curves of the filters. By comparison with Fig.~\ref{h2hj_zoom_ysos} where 
the positions of YSOs are overlaid, we see that the Br$\gamma$ emission peaks approximately 5$\arcsec$ 
to the south of Star $A$. There is also a rim of Br$\gamma$ emission along the northern edge of the 
globule, which we interpret as being due to the external UV field from the Cygnus OB2 cluster, located 
to the North of the globule (see Fig.~\ref{overview}). This emission is much fainter than the centrally 
peaked emission, however, which we argue is due to the internal feedback from the stars inside the 
globule. The highest intensity is found around the binary Class~I star $C$ and coincides with the 
brightest peak of \OI\ emission \citep[see][Fig.~3]{sch12}, thus highlighting the extreme youth 
of this part of the globule.

The comparison between the high-angular-resolution images tracing the PDR and the ionised gas 
emission highlights the striking differences between their respective distribution, already 
hinted in \citet{sch12}. In the H$_2$ images, the PDRs appear mainly as bright fingers of emission 
highlighting rims where the molecular gas is being photodissociated. Figure~\ref{nb_rgb_im} 
strongly suggests that the molecular gas traced by the PDR emission is approximately distributed in 
a shell that delineates the outer contour of the globule. The globule and its tail is initially 
shaped by external feedback from the UV field with the globule head pointing towards the OB2 
cluster with the tail away from it. The brightness pattern of the emission rims, however, suggests 
that they are illuminated from the inside, reinforcing the interpretation of \citet{sch12} in 
favour of an internal origin for the PDR emission caused by the embedded cluster, rather than an 
external origin due to the UV field. On the other hand, the clear central condensation of the 
Br$\gamma$ emission, together with the spectroscopic evidence that we present in 
Sect.~\ref{neb_spect}, clearly show that the ionised gas is also caused by the embedded cluster 
stars.

\subsubsection{Spectroscopy of the nebula \label{neb_spect}}

The extracted optical spectrum of the nebula approximately 9$''$ East of the position of star~$A$ 
clearly indicates a low-excitation \HII\ region. No hints of [O\,{\scriptsize III}] emission are 
seen near 5000~\AA, {and there is only a marginal detection of He\,{\scriptsize I} at 6678~\AA , 
with He\,{\scriptsize I} at 5875~\AA\  being undetectable (Fig.~\ref{nebula_spec_vis}). This is what 
one would expect from a small \HII\ region ionised by the soft radiation of early B-type stars in 
the globule, rather than by the harder 
radiation field dominated by the O stars of nearby Cygnus~OB2. This leads us to conclude that the 
radiation field inside the globule is dominated by its embedded B stars, rather than by harder 
radiation leaking in from the outside. The electron density inferred from the [S\,{\scriptsize II}]
 lines is $n_e \simeq 200$~cm$^{-3}$.

\begin{figure}[ht]
\includegraphics[angle=0,width=9cm]{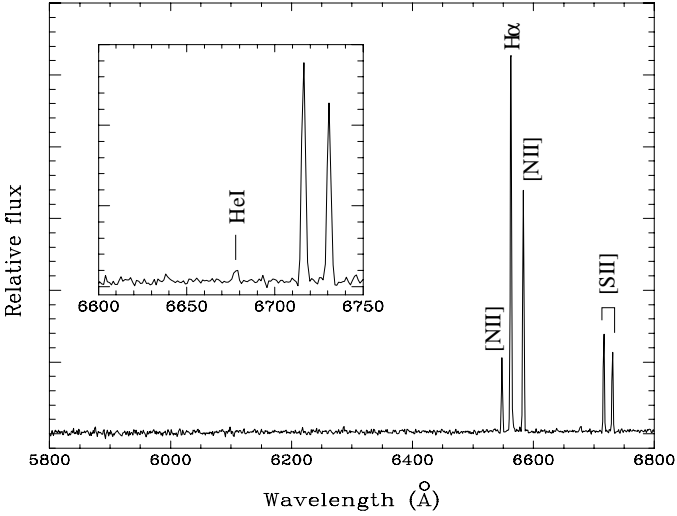}
\caption [] {Visible-red spectrum of the \HII\ region. The intensity ratio of [S\,{\scriptsize II}] 
lines at 6717 and 6731 indicates an electron density $n_e \simeq 200$~cm$^{-3}$. The marginal 
detection of He\,{\scriptsize I} at 6678~\AA \ (see inset) and the non-detection of 
He\,{\scriptsize I} at 5875~\AA\ indicate a soft ionising radiation like that expected from the early 
B stars embedded in the globule.}
\label{nebula_spec_vis}
\end{figure}

\section{Discussion}

Our observations reveal a cluster of very young stellar objects associated with the globule
IRAS~20319+3958. Youth is clearly highlighted by the ratio of Class~I to Class~II sources,
implying that star formation there has been proceeding for a period comparable to the typical
span of the Class~I stage. Signposts of this youth are also found among the brightest young
stellar objects of the globule: one of them (star~$A$) is a binary in which at least one of the
components is a Herbig Be star, and the other two, star~$B$ and the binary Class~I star~$C$,
have very compact, low-excitation \HII\ regions around them. An obvious upper limit to the number
of diskless (Class~III) young stellar objects is given by the number of stars that appear
projected on the area of the globule and that do not display infrared excess, as compared
to the number of those same sources in a typical off-cloud area. We find no significant
enhancement of non-excess sources at the position of the globule as compared to the background
population, indicating that the number of Class~III sources is small and probably negligible
in comparison to that of Class~I and II sources.

Some extreme Class~I or even Class~0 objects surrounded by thick envelopes may be missed by
the near-infrared observations, but the lack of sources detected by {\it Spitzer} and without a
near-infrared counterpart in our $JHK_S$ images argues against this being a numerically
significant population. We thus conclude that, although having taken place very recently,
the star-formation process in IRAS~20319+3958 is essentially finished now.
  The high Class~I-to-Class~II ratio together with the absence of a significant Class~III
population suggests that the age of the cluster is much shorter than typical disk dispersal
timescales. The latter are suspected to depend on the central object's mass, the stellar
density of the environment \citep{luh10} and binarity \citep{dae16}, but disk lifetimes of
1-3~Myr appear to be typical, both on a theoretical and an observational basis 
\citep[e.g.,][]{lix16}. The duration of the Class~I phase has been traditionally estimated by 
comparing the relative number of Class~I to Class~II YSOs in star-forming regions under the 
assumption of a constant star-formation rate \citep{eva09}, but recent work based on detailed 
observations and modelling of the accretion on the central object have challenged an 
evolutionary picture in which the circumstellar envelope steadily disperses until the 
Class~I/II transition takes place, suggesting instead that Class~I objects may reach ages 
comparable to those of T~Tauri stars, up to 1~Myr and older \citep[][and references therein]{whi07}. 
We can thus take 1~Myr as a 
reasonable estimate of the age of the globule population, as already anticipated in 
Section~\ref{phys_ysos}. However, if the more traditional interpretation of Class~I objects as 
a population distinctly less evolved and shorter lived than the Class~II population is adopted, 
then the Class~I/Class~II ratio argues for an even younger age of the aggregate.

The youth of the cluster leads us to wonder if its formation could have been triggered by the 
passage of the ionisation front that left it as an isolated feature. Evidence for triggered star 
formation near rim nebulae and clumps exposed to eroding ultraviolet radiation, although elusive, 
has been reported in a number of cases
\citep[e.g.][]{sic14,get07,niw09,hay12,com05,com07}
under conditions that closely resemble those found in IRAS~20319+3958. {\it Spitzer} and molecular 
line images of the region (see Fig.~\ref{overview} and \citet{sch16}) show the abrupt interface 
between the Cygnus X molecular complex and the interstellar medium cleared by the ionisation and 
winds from the stars in Cygnus OB2. In the region where the globule lies, such interface, which 
marks the current location of the ionisation front, is located approximately 24$'$ South of the 
globule, corresponding to a projected distance of 10.5~pc (see Fig.~\ref{overview}). If the globule 
had started forming stars spontaneously before being run over by the ionisation front, a cluster 
age of 1~Myr would require an averaged propagation speed of the ionisation front of approximately
10~km~s$^{-1}$ or more. The possibility that star formation started before the arrival of the 
ionisation front remains plausible in view of our results, unless our choice of 1~Myr as the age 
of the aggregate grossly overestimates its actual age, which is unlikely in view of its sizeable 
Class~II population. Our results, therefore, do not unambiguously argue for the passage of the 
ionisation front as the factor that triggered star formation in the globule. However, the approximate 
coincidence of the onset of star formation in the globule with the estimated passage of the 
ionisation front, if the latter propagated at 10~km~s$^{-1}$ \citep[see][]{sch16},
is circumstantial evidence for a causal connection between both events. We note in support of this 
connection that the distribution of Class~II sources in the periphery of the globule, preferentially 
projected on the side facing the densest regions of the Cygnus~OB2 as noted in Sect.~\ref{yso_sample}, 
also suggests that the changes in the environment of the globule caused by its 
rapid exposure to the ionising radiation \citep{ber89,gri09,tre12} played a determinant role in 
the star-formation history of the globule.

The hints of an atypical mass function described in Sect.~\ref{phys_ysos}, if real, might also be
related to the particularities of star formation in a globule exposed to intense external ultraviolet
radiation. Our results resemble those obtained on another star-forming region suspected to be externally
irradiated, \object{IRAS~16362-4845} in \object{RCW108}, where we also found indications of a top-heavy
mass function \citep{com07}. However, possible departures from a universal form of the initial mass
function are far from established. Examining the excess of young stellar objects around bubbles as
proof of triggered star formation, \citet{tho12} find no evidence for an initial mass function different
from that in the field. Similarly, in their detailed study of the stellar population in the Trumpler~37
cluster and its adjacent globule IC~1396A, \citet{get12} obtain a good fit to their results with the
widely used initial mass function of \citet{kro01}, for the populations of both the central cluster and 
the globule, although the latter does not contain stars more massive than 2~M$_\odot$. On the theoretical
side, the problems in relating the effects of temperature, turbulence, and magnetic fields with the
peak of the mass function and its slope at different masses have been outlined by \citet{off14},
but no clear connection between peculiarities in the initial mass function and the star-formation
conditions expected as the result of the triggering mechanism has been demonstrated.

\section{Summary and conclusions}

We studied the content of the cluster embedded in the IRAS~20319+3958 globule and the distribution 
of its excited molecular hydrogen and ionised gas by means of visible and near-infrared imaging and 
spectroscopy, complemented with mid-infrared {\it Spitzer}/IRAC imaging. The study complements that of 
\citet{sch12} on the physical and kinematical conditions of the molecular and photodissociating gas, 
thus providing a comprehensive picture of the globule and its content. Our conclusions can be summarised 
as follows:

\begin{itemize}

\item The IRAS~20319+3958 globule contains an embedded aggregate of approximately
  30 very young stellar objects, most of them characterised by
  the strong near and/or mid-infrared excesses expected from a very
  early evolutionary stage. We estimate the age of the embedded
  aggregate to be $\sim 1$~Myr, based on the high observed ratio of
  Class~I to Class II objects. We also estimate a stellar mass of the
  aggregate of $\sim 90$~M$_\odot$.

\item The most massive members of the aggregate are three systems
  containing early B-type stars. Two of them are able to produce very
  compact \HII\ regions, one of them being still highly embedded and
  coinciding with a peak in PDR emission. Two of these three
  systems are resolved binaries, and one of them contains a visible
  Herbig Be star.

\item The mass function of the aggregate, based on an approximate derivation
  of the individual stellar masses, suggests a high-mass tail characterised
  by a slope shallower than that found in most clusters, and a peak of the
  mass function possibly lying at higher masses. This might be due to the
  influence of the strong ionising radiation field under which the aggregate
  was formed. However, large uncertainties affecting our derivation of the 
  individual masses of the members of the aggregate, most notably the age 
  adopted for its faintest members, make this conclusion tentative at best.

\item The compared morphologies of the H$_2$ emission that traces the
  photon-dominated regions and the Br$\gamma$ emission that traces the
  ionised gas suggest that molecular gas is roughly distributed as a
  shell around the embedded aggregate, filled with centrally-condensed
  ionised gas. Both the morphology and the low excitation of the \HII\
  region indicate that the sources of ionisation are the B stars of
  the embedded aggregate, rather than the external UV field caused by
  the O-stars of Cygnus~OB2.

\item Considering the estimated age of the embedded aggregate and the
  isolation of the globule, we find it likely that the formation of the embedded 
  aggregate started when
  the globule was overtaken by the large-scale ionisation fronts that
  the O stars of Cygnus~OB2 drive into the Cygnus~X region.

\end{itemize}

The {\it Spitzer} images of the Cygnus~OB2 / Cygnus~X region offer a dramatic view
of pillars and globules resulting from erosion by the massive association of its 
parental molecular gas, and, simultaneously, of the recent star formation revealed by 
the strong infrared excesses caused by hot circumstellar dust. Those images strongly 
suggest that the globule IRAS~20319+3958 is a typical feature in the region rather 
than a rarity, and the same is apparent from the images of other massive star-forming 
regions. The study of globules and pillars and their embedded populations can thus 
provide excellent samples for the study of triggered star formation and the features
that it may imprint on the stellar aggregates resulting from it.

\begin{acknowledgements}

Based on observations made with Omega2000 at the 3.5m telescope of the Centro
Astron\'{o}mico Hispano Alem\'{a}n (CAHA) at Calar Alto, operated jointly by the
Max-Planck Institut f\"{u}r Astronomie and the Instituto de Astrof\'{i}sica de
Andaluc\'{i}a (CSIC). Partly based on observations made with the Nordic Optical Telescope,
operated by the Nordic Optical Telescope Scientific Association at the Observatorio del
Roque de los Muchachos, La Palma, Spain, of the Instituto de Astrof\'{i}sica de Canarias,
as well as observations with the IAC80 telescope, operated on Tenerife by the
Instituto de Astrof\'{i}sica de Canarias in the Spanish Observatorio del Teide.
NS acknowledges support by the ANR-11-BS56-010 project STARFICH and support from the 
Deutsche Forschungsgemeinschaft, DFG, through project number Os 177/2-1 and 177/2-2, 
central funds of the DFG-priority program 1573 (ISM-SPP), and the SFB 956. We thank the 
Calar Alto staff for the execution of our observations in Service Mode. We thank Stefan 
Geier, Jussi Harmanen, and Joonas Sario for some of the observations during their 
studentships at the NOT.
\end{acknowledgements}


\bibliography{globule_ref}


\listofobjects

\end{document}